\documentclass[aps,prb,twocolumn,superscriptaddress,showpacs,showkeys]{revtex4-2}
\usepackage[pdftex]{graphicx}
\usepackage{color}
\usepackage{amsmath}
\usepackage{multirow}

\usepackage{changes}

\usepackage{xr}
\begin{document}


\title{Multiplet Effects in the Electronic Correlation 
of One-Dimensional Magnetic Transition-Metal Oxides on Metals}



\newcommand{\QUIMI}[0]{{
Departamento de Pol\'{\i}meros y Materiales Avanzados: 
F\'{\i}sica, Qu\'{\i}mica y Tecnolog\'{\i}a, Facultad de Qu\'{\i}mica UPV/EHU,
Apartado 1072, 20080 Donostia-San Sebasti\'an, Spain}}

\newcommand{\DIPC}[0]{{
Donostia International Physics Center, 
Paseo Manuel de Lardiz\'abal 4, 20018 Donostia-San Sebasti\'an, Spain}}

\newcommand{\CFM}[0]{{
Centro de F\'{\i}sica de Materiales CFM/MPC (CSIC-UPV/EHU), 
Paseo Manuel de Lardiz\'abal 5, 20018 Donostia-San Sebasti\'an, Spain}}

\newcommand{\JUELICH}[0]{{
Peter Gr\"unberg Institut and Institute for Advanced Simulation, 
Forschungszentrum J\"ulich and JARA, 52425 J\"ulich, Germany}}

\author{J. Goikoetxea}
\affiliation{\CFM}
\author{C. Friedrich}
\affiliation{\JUELICH}
\author{G. Bihlmayer}
\affiliation{\JUELICH}
\author{S. Bl\"ugel}
\affiliation{\JUELICH}
\author{A. Arnau}
\affiliation{\CFM}
\affiliation{\QUIMI}
\affiliation{\DIPC}
\author{M. Blanco-Rey}
\affiliation{\QUIMI}
\affiliation{\DIPC}



\date{\today}

\begin{abstract}

We use the constrained random phase approximation (cRPA) 
method to calculate the Hubbard $U$ parameter
in four one-dimensional magnetic transition metal atom oxides 
of composition XO$_2$ (X = Mn, Fe, Co, Ni) on Ir(100).  
In addition to the expected screening of the oxide, i.e., 
a significant reduction of the $U$ value by the presence of the metal substrate, 
we find a strong dependence on the electronic configuration (multiplet) 
of the X($d$) orbital. Each particular electronic configuration attained by atom X 
is dictated by the O ligands, as well as by the charge transfer and 
hybridization with the Ir(100) substrate. We find that MnO$_2$ and NiO$_2$ 
chains exhibit two different screening regimes, while the case of CoO$_2$ 
is somewhere in between. The electronic structure of the MnO$_2$ chain 
remains almost unchanged upon adsorption.
Therefore, in this regime, the additional screening is predominantly generated 
by the electrons of the neighboring metal surface atoms.
The screening strength for NiO$_2$/Ir(100) is 
found to depend on the Ni($d$) configuration in the adsorbed state. 
The case of FeO$_2$ shows an exceptional behavior, as
it is the only insulating system in the absence of metallic substrate and, thus, 
it has the largest $U$ value. However, this value is significantly reduced by the 
two mentioned screening effects after adsorption.

\end{abstract}


\maketitle

\section{Introduction}
\label{sec:intro}

The metallic or insulating character of a system often lies
beyond the descriptive power of one-electron models. 
In systems with localized electrons, such as in $d$-electrons,  
the existence of a gap is determined by three quantities:
(i) the charge transfer (CT) gap $\Delta_{Ld}$ from a ligand ($L$) to the open $d$ shell,
(ii) the bandwidth that results from hybridization, and
(iii) the $d$-orbital intrashell electron-electron Coulomb interaction, 
usually expressed in terms of correlation $U$ and exchange $J$ parameters. 
The renowned Zaanen-Sawatzky-Allen (ZSA) diagram abridges the various 
insulating and metallic phases resulting from these interactions \cite{bib:zaanen85}.
Samewise, exchange interactions between spin-polarized ions stem
from the aforementioned terms of the Hubbard Hamiltonian, 
as a direct consequence of the competition between electron itinerance and 
intrashell Coulomb interaction.
While the bare interaction amounts to a few tens of eV in $3d$ 
transition metals (TM), the effective interaction is strongly damped 
by the electronic screening. Therefore, its magnitude depends on the 
electronic structure of the particular system, e.g., on the hybridization 
of the $d$ orbitals and charge transfer effects. 
It has been found that the effective interaction parameter
is of a few eV in bulk oxides and, at the 
metal/oxide interface, the Coulomb interaction is further screened 
by twice the image charge formation potential \cite{bib:duffy83,bib:altieri99}. 
The dimensionality $d$ determines the screening length:
for $d \le 2$, long-range screening is suppressed and anti-screening may exist 
at intermediate inter-atomic distances \cite{bib:vandenbrink00,bib:peters17}.

Several electronic structure methods rely on realistic effective $U$ parameters
to account for screening mechanisms as an alternative to more complex calculations.
In this context, the popular LDA+$U$ \cite{bib:anisimov97} used in density functional theory (DFT)
allows to include correlation effects via an orbital-dependent functional 
that is applied specifically to the localized orbitals.
It can be formulated as a rotationally invariant functional
dependent on constant $U$ and $J$ values \cite{bib:liechtenstein95}.
In the next level of complexity we find many-body methods beyond the 
one-electron picture of DFT, for example, 
LDA++ \cite{bib:lichtenstein98}, 
GW \cite{bib:hedin65}, 
MP2 \cite{bib:moller34} or RPA \cite{bib:langreth75} and, notably,
methods based on dynamical mean-field theory (DMFT) \cite{bib:kotliar06,bib:biermann14,bib:biermann03}.

LDA+$U$ and DMFT-based methods require suitable $U$ and $J$ 
parameters to predict correlation-dependent properties.
A number of methods have been proposed to determine those quantities self-consistently 
from first principles, such as constrained DFT \cite{bib:dederichs84,bib:cococcioni05}
(recently improved to avoid supercell calculations \cite{bib:timrov18,bib:timrov21})
and constrained random-phase approximation (cRPA) \cite{bib:arya04,bib:arya06}.
The latter approach also gives access to the frequency dependence of the interaction.
In it, the polarizability contribution of electrons in the correlated subspace
is excluded to obtain a screened interaction $\hat U$, whose matrix elements in the 
localized basis are the sought-for $U_{mn,m'n'}$ 
Coulomb matrix elements of the Hubbard Hamiltonian, 
where indices label the orbitals.
In practice, maximally localized Wannier functions (MLWF) are used as basis sets
\cite{bib:pizzi20,bib:marzari97,bib:miyake09,bib:sasioglu11}. 


In this paper, we use cRPA to calculate $\hat U$ in four one-dimensional
magnetic transition-metal oxides (TMO) 
of composition XO$_2$ (X = Mn, Fe, Co, Ni) on Ir(100).
In addition to the expected screening effect  
of the metal substrate, we find a strong dependence of the $U$ value on the electronic 
configuration (multiplet) of the X($d$) orbital. 
The particular configuration attained by X is dictated by the O ligands,
as well as by the charge transfer and hybridization with the substrate. 
MnO$_2$ and NiO$_2$ chains represent two different regimes. 
The electronic structure of the MnO$_2$ chain remains 
almost unchanged upon adsorption.
In this regime screening by the neighboring  metal surface atoms applies.
In contrast, NiO$_2$/Ir is in the other regime, where screening is dominated by the Ni($d$) 
configuration adopted in the adsorbed state.
The cRPA calculations show that the multiplet effect cannot be uncoupled 
from screening by the metal. 

XO$_2$ chains grown in ultra-high vacuum on Ir(100) are aligned along the 
$[100]$ crystallographic direction and can reach lengths up to 130\,nm \cite{bib:ferstl16}.  
The chains self-organize in a $(3 \times 1)$ missing-row superstructure (see Fig.~\ref{fig:bands_nio2}(c)) with 
rotational domains of $\sim 100$\,nm$^2$ extension.
Low-energy electron diffraction [LEED-I(V)] 
shows that the Ir atoms below the chain are lifted to leave room 
for the X atoms, which are not coplanar with the oxygens \cite{bib:ferstl16,bib:ferstl17}. 
Pt(100) can serve as growth template as well \cite{bib:ferstl17,bib:schmitt19b}.
DFT+$U$ calculations in the literature have used a low value $U-J=1.5$\,eV for  
these systems on the premise that the interactions within the X($d$) orbital 
are heavily screened by the metal substrate \cite{bib:ferstl16,bib:schmitt19b}. Weak 
antiferromagnetic (AFM) coupling is found along 
MnO$_2$ and CoO$_2$ on Ir(100) \cite{bib:ferstl16} and CoO$_2$ on Rh(553) \cite{bib:korobova20}, 
FeO$_2$/Ir(100) is ferromagnetic (FM) and the Ni spin moment 
in NiO$_2$/Ir(100) is quenched \cite{bib:ferstl16}. 
A long-range chiral non-collinear modulation is also observed along MnO$_2$/Pt(100)
using scanning tunneling microscopy \cite{bib:schmitt19b}.
Substrate-mediated RKKY exchange promotes additional lateral interactions 
between the MnO$_2$/Ir(100) chains, which are also non-collinear and chiral \cite{bib:schmitt19}.
The modelling of these magnetic properties is subject to understanding Coulomb interactions.
As a matter of fact, a variation in the Curie and N\'eel critical temperatures 
has been observed during magnetic oxide film growth on metals 
that can be explained by the image potential screening length \cite{bib:altendorf18,bib:barman20}.
As a first approximation, the magnetic exchange along the chain 
follows from the electron hopping through X($d$)-O($p$)-X($d$) orbitals 
(superexchange) \cite{bib:goodenough55,bib:kanamori59}.
In the CT insulator limit ($U \gg \Delta_{pd}$) \cite{bib:zaanen85}, 
the coupling constant of the AFM channel roughly scales as 
$\sim t^4/U^3$, where $t$ accounts for the hopping integrals, and the FM one as $\sim -t^2/U$.
Additionally, the four-fold coordination of X($d$) allows for competing hopping pathways 
that tend to weaken the magnetic interactions \cite{bib:korobova20}. 
All in all, the combined effect of the O($p$) ligands and Ir substrate screening channels
will lead to a system-dependent renormalization of intraorbital X($d$) interactions 
with consequences for the previously described magnetic properties.
This understanding of system dependent interactions is the purpose of this study.

The paper is organized as follows: in Section~\ref{sec:theory} we describe the 
DFT and cRPA calculations; the results and discussion are presented in 
Section~\ref{sec:results}, which is divided into Subsections~\ref{sec:isolchains} 
and \ref{sec:irchains} on the isolated and the adsorbed chains on Ir(100), respectively. 
Finally, conclusions are drawn in Section~\ref{sec:conclusions}.

\section{Theoretical Methods}
\label{sec:theory}

\begin{figure}[tb!]
\centerline{\includegraphics[width=1\columnwidth]{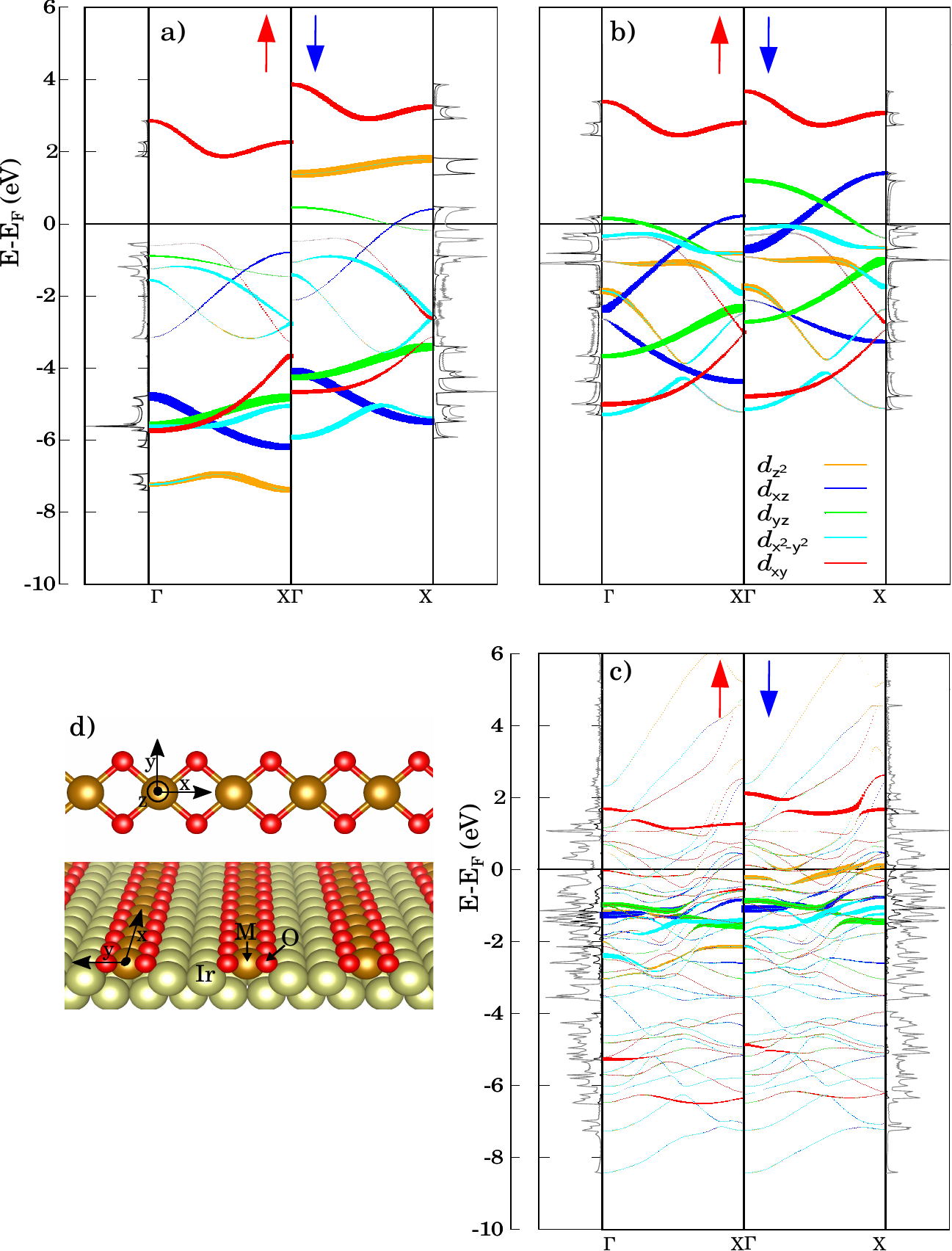}}
\caption{
Band structures (central subpanels)
and densities of states (lateral subpanels) for NiO$_2$ chains,
free-standing in the C1 (panel a) and C2 (b) configurations and supported on Ir(100) (c).
Spin majority and minority contributions are indicated by red and
blue arrows, respectively. The colour code indicates the Ni($d$) orbital resolved bands,
obtained from projection on the corresponding MLWFs,
with dot sizes accounting for the magnitude of the projection.
The full bands are shown as gray dotted curves for the isolated chains only.
The solid gray and black curves are the total and Ni($d$)-projected densities of states,
respectively, in each system in arbitrary units (the gray curve in the bottom panel does
include the Ir atoms contribution). 
The model (d) shows the atomic structure of the free-standing and adsorbed chains
on the missing row $(3\times1)$ Ir(100) surface.}
\label{fig:bands_nio2}
\end{figure}

Ideal free-standing planar XO$_2$ chains are modelled in the supercell approach.
The geometry is found by relaxing the X-X distances and X-O bonds 
in a calculation with the \textsc{VASP} code (projector augmented wave potentials 
and a plane-wave basis set \cite{bib:kresse93,bib:kresse99}) 
using the GGA+$U$ approximation \cite{bib:ferstl16,bib:schmitt19} .
The details of the geometry determination are described in the Supplementary Material (SM) 
Tables~\ref{SM-tab:struc_free} and \ref{SM-tab:struc_adsorbed}.
For the XO$_2$/Ir(100) model structures (see Fig.~\ref{fig:bands_nio2}(d)), 
a $(3\times 1)$ missing-row reconstructed substrate with the 
experimental in-plane lattice constant $a = 2.71$\,{\AA} is used, 
as found in the LEED-I(V) study of Ref.~\cite{bib:ferstl16}.
The substrate slab consists of five monolayers, where the bottom 
layer is kept fixed during the relaxation. 
Two layers are kept in the cRPA calculations 
\footnote{Inclusion of a third Ir layer in a cRPA iteration
changes the obtained $U$ value for MnO$_2$/Ir by $\sim 10$\%. }. 
These geometry relaxations have been carried out at a fixed 
value $U=1.5$\,eV, used also in other works with supported 
XO$_2$ chains \cite{bib:ferstl16,bib:schmitt19}.
In the coplanar isolated chains 
the equilibrium geometry shows low sensitivity to the $U$ value,
but the spin state can be altered by a change in $U$ (see SM Fig.~\ref{SM-fig:spin_phases}).

The one-electron wavefunctions used in the cRPA calculations are obtained 
from {\it ab-initio} DFT+$U$ calculations \cite{bib:anisimov97}
with the generalized-gradient PBE functional \cite{bib:pbe96} 
in the fully-localized limit \cite{bib:anisimov93} 
for the LDA+$U$ double-counting term. 
The used code, \textsc{Fleur} \cite{bib:fleur}, is based on the 
full-potential linearized augmented plane waves (FLAPW)
formalism \cite{bib:krakauer79,bib:wimmer81,bib:shick99}.  
The local basis functions are expanded to $l_{max}=8,6$ and 8 inside  
the X, O and Ir muffin tin spheres, of radii 2.1, 1.2 and 2.4\,a.u., respectively.
The wavevector cut-off in the interstitial region is 5\,a.u.$^{-1}$ 
for wavefunctions and 14\,a.u.$^{-1}$ for the potential, using 
a $10 \times 3 \times 3$ Brillouin zone sampling
(the sampling in the perpendicular direction is needed for constructing the 
Wannier functions).
The Fermi level was determined by a Fermi-Dirac smearing of width 0.27\,eV.
To adapt these parameters to each particular case, slight modifications 
are performed that can be found in the SM Table~\ref{SM-tab:flapw_params}.

For the unsupported XO$_2$ chains, MLWFs \cite{bib:pizzi20,bib:marzari97} 
for 11 bands are constructed with projections 
of $d$ type on X=Fe,Co,Ni and $s,p_x,p_z$ or $p=(p_x,p_y,p_z)$ orbitals on O. 
For MnO$_2$, 13 bands are needed ($d$ on Mn and $s,p$ on O). 
In the supported case, the Wannier basis needs to
be extended to include the Ir states, since the 
oxide states are strongly hybridized with the metal substrate.
Up to 41 bands are considered, including $p,d$ projections on X, 
$s,p$ on O and $sp^3d$ on Ir. 
The static partially screened Coulomb interaction matrices in the Wannier basis 
are calculated with the cRPA formalism as implemented in the code
\textsc{Spex} \cite{bib:friedrich10},
where correlated subspaces can be selected {\it ad hoc} to obtain
information about different screening channels in the system.
The effective parameters $U,J$ are obtained by averaging the Coulomb interaction 
matrix elements as described in Ref.~\cite{bib:anisimov93}.

In cRPA, the Hilbert space is divided into the correlated ($d$) 
and the remainder ($r$) parts. The two subspaces are orthogonal. 
The total polarization is $\hat P = \hat P_d + \hat P_r$, where $P_d$ contains 
$d \rightarrow d$ transitions only, while $\hat P_r$ accounts for 
$d \rightarrow r$, $r \rightarrow d$ and $r \rightarrow r$ transitions.
The fundamental equations of the cRPA formalism \cite{bib:arya04} are 
(dropping the frequency dependence) 
$\hat W=(1-\hat W_r \hat P_d)^{-1} \hat W_r$ and 
$\hat W_r=(1-\hat v \hat P_r)^{-1} \hat v$, 
where $\hat v$, $\hat W_r$ and $\hat W$ are the bare, 
effective and fully-screened Coulomb 
interactions, respectively. The spherical average of the static limit of $\hat W_r$ 
is the sought-for Hubbard $U$ parameter.
The use of MLWFs basis to isolate the localized states belonging to the $d$ 
subspace is commonplace and there are different methods to uncouple the $r$ and $d$ 
subspaces \cite{bib:miyake09}. 
Here, we use the one described in Ref.~\cite{bib:sasioglu11}, which is fully 
basis-set-independent. In it, the transitions are weighted by the probability 
that the initial and final states belong to the $d$ subspace. 
Finally, the effective Coulomb interaction matrix elements $U_{mn,m'n'}$ are calculated
for a selected subset of functions out of the whole Wannier set.
In order to interpret the $U_{mn,m'n'}$ matrix elements obtained in this
cRPA approach as atomic Coulomb integrals, we make sure that the MLWFs conserve the 
symmetry of the atomic orbitals. The MLWFs are also used to calculate 
band structure projections and orbital occupancies.


\section{Results and discussion}
\label{sec:results}

\subsection{Isolated planar XO$_2$ chains}
\label{sec:isolchains}

In this subsection, we study an ideal situation where  the 
main substrate effects are suppressed, 
namely the geometry change (the here-assumed coplanarity between the X and O atoms
is lost in the adsorbed geometry) and charge redistribution
at the metal-oxide interface.
These idealized planar chains 
allow us to examine the screening originated exclusively by the 
X-O bond formation and the one-dimensionality of the system.

All the cRPA calculations are initialized with the electronic structures 
obtained with $U_0=5.5$\,eV and $J_0=0$. 
Once the first $U_1$ is calculated using cRPA, a new DFT+$U$ cycle is started with this value.
After 4-5 iterations, convergence is achieved for MnO$_2$,
FeO$_2$ and CoO$_2$. The resulting $U$ and $J$ values for the $\uparrow\uparrow$ spin channel, 
shown in Table~\ref{tab:UJ}, lie in the 5-7\,eV range and 
are obtained independently of the $U$ value used in the initial iteration 
(this was checked by starting from $U_0=3.5$ and 7.5\,eV). 
The effective parameter $U$ plays the role of the $F^0$ Slater 
integral in the Coulomb interaction term and the effective 
intra-atomic exchange parameter is $J=(F^2+F^4)/14$ in the case of a $d$-orbital \cite{bib:anisimov93}. 
$F^2$ and $F^4$ are known to be almost 
insensitive to screening effects \cite{bib:vandermarel88,bib:altieri99} and,
in fact, we obtain the typical value $J\simeq 1$\,eV in the cRPA calculations.
Due to the spin dependence of the single-particle states, 
the Wannier functions exhibit a spin dependence, too. 
As a consequence, we can distinguish between the matrix 
elements $U^{\uparrow\uparrow}$, $U^{\uparrow\downarrow}$, 
and $U^{\downarrow\downarrow}$. They exhibit a spread of $\approx$0.5\,eV 
about their average values for FeO$_2$ and MnO$_2$,
and of $\approx 0.2$\,eV for NiO$_2$ and CoO$_2$ 
(see SM Table~\ref{SM-tab:Uspinchannels}).
Spreads in the $J$ values are $\le 0.2$\,eV.

\begin{table}[tb!]
\caption{\label{tab:UJ} Converged values of $U$ and $J$, averaged over the $\uparrow\uparrow$ spin channel orbitals,
for the transition metal atoms in free-standing planar and Ir-supported XO$_2$ chains.
$\tilde U$ are the values calculated in the shell folding approach \cite{bib:seth17}.
All units in eV.}
\begin{ruledtabular}
 \begin{tabular}{c  c  c  c | c  c  c}
   XO$_2$  &  $U$  & $J$  & $\tilde U$ &   XO$_2$/Ir  &  $U$ & $J$ \\
   \hline
   Ni (C1) &  6.59 & 1.17 & 8.45 &  \multirow{2}{*}{Ni}     &   \multirow{2}{*}{1.71} & \multirow{2}{*}{0.87} \\
   Ni (C2) &  2.41 & 1.01 & 7.03 &    &      &       \\
   Mn      &  6.21 & 1.04 & 6.57 & Mn & 3.78 & 0.98  \\
   Co      &  5.73 & 1.11 & 8.62 & Co & 2.39 & 0.90  \\
   Fe      &  7.67 & 1.13 & 9.06 & Fe & 1.38 & 0.80
 \end{tabular}
\end{ruledtabular}
\end{table}

The case for NiO$_2$ deserves further attention.
Unlike in the other chains, two electronic configurations of the Ni($d$) orbital 
are stabilized in the initial run for different $U_0$ values, 
labelled C1 and C2 hereafter. 
The C1 and C2 configurations are preserved
throughout the subsequent cRPA cycles, which converge to two different
$U$ values (see Table~\ref{tab:UJ}).
DFT+$U$ calculations with $U_0 < 4$\, eV yield C2 as the most stable
configuration and finally $U=2.41$\,eV, while C1 is the preferred configuration
for $U_0 \ge 4$\, eV, leading to $U=6.59$\,eV.
The details of $d$-orbital occupancies that define C1 and C2 
are gathered in Table~\ref{tab:occup_nio2} for the DFT+$U$ 
calculations at the converged $U$ values.
Occupancies are calculated as integrals of the projected densities of states (PDOS) 
on the individual $d$-like MLWFs (see SM Fig.~\ref{SM-fig:pdos_nio2}). 
The projected band structure is shown in Fig.~\ref{fig:bands_nio2}. 

\begin{table}[tb!]
\caption{\label{tab:occup_nio2} $d$-orbital occupancies and magnetic spin moments of the Ni atom
at free-standing (C1 and C2 configurations) and Ir-supported NiO$_2$. }
\begin{ruledtabular}
\begin{tabular}{c  c  c  c  c  c  c  c}
  & Spin & $d_{z^2}$ & $d_{xz}$ & $d_{yz}$ & $d_{x^2-y^2}$ & $d_{xy}$ &   $\mu_{\text{Ni}}\; (\mu_B)$ \\
   \hline
  \multirow{2}{*}{NiO$_2$: C1} & $\uparrow$    & 0.99   & 0.99 & 0.99 & 0.99 &  0.54  & \multirow{2}{*}{1.23} \\
   &  $\downarrow$  & 0.03   & 0.94  &  0.85 & 0.97 & 0.42 &\\
  \hline
 \multirow{2}{*}{NiO$_2$: C2} & $\uparrow$    &  0.99 & 0.91 & 0.86 & 0.99 & 0.48 & \multirow{2}{*}{0.55} \\
 & $\downarrow$  &  0.99  & 0.64 & 0.61 & 0.99 & 0.46 \\
  \hline
 \multirow{2}{*}{NiO$_2$/Ir} & $\uparrow$    & 0.97 & 0.95 & 0.95 & 0.97 & 0.62 & \multirow{2}{*}{0.52}\\
 & $\downarrow$  & 0.59 & 0.94 & 0.95 & 0.96 & 0.41
\end{tabular}
\end{ruledtabular}
\end{table}

\begin{figure}[tb!]
\centerline{\includegraphics[width=1\columnwidth]{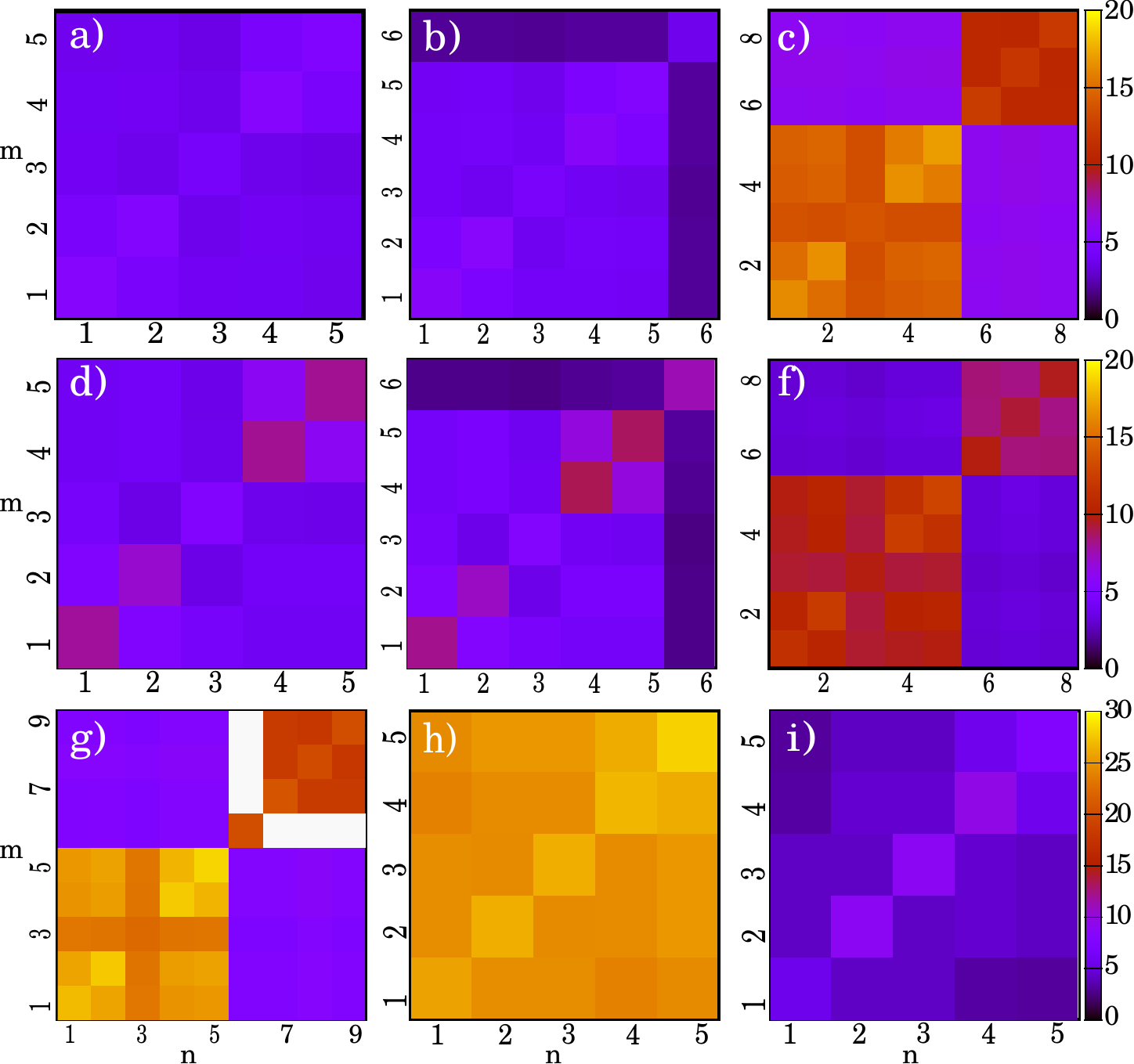}}
\caption{
Screened Coulomb matrix elements $U_{mn}$ 
($\uparrow \uparrow$ spin channel) in eV for the NiO$_2$ chain for
selected orbitals screened by the remaining electrons.
For the free-standing C1 configuration, the selection is: (a) $d$
(indices ordered as $1-d_{z^2}$, $2-d_{xz}$, $3-d_{yz}$, $4-d_{x^2-y^2}$ and $5-d_{xy}$),
(b) $d$ and O($s$) (index 6)  and (c) $d$ and O($p$) orbitals
(indices ordered as $6-p_z$, $7-p_x$ and $8-p_y$).
(d-e) shows the same information as (a-c) but for the free-standing C2 configuration.
In the bottom row, bare Coulomb matrix elements of
(g) free-standing C1 and (h) Ir-supported NiO$_2$,
which we take as reference values.
(i) For Ir-supported NiO$_2$, $U_{mn}$ elements of the $d$
orbital screened by the rest of electrons. Note the different color scale for
the panels (g) to (i)}
\label{fig:check_nio2}
\end{figure}

The Ni spin moments obtained for the C1 and C2 configurations, 1.23 and 0.55\,$\mu_B$, 
deviate from an integer value. The closest integer is 1\,$\mu_B$ in 
both cases, which leads us to interpret C1 and C2 to be
two Ni multiplets of the same {\it nominal} spin state $S=1/2$.
The bare Coulomb parameters 
in the C1 and C2 configurations take essentially similar values, 
differences between individual matrix elements being, on average, 8\%.
This small difference in the bare Coulomb matrix is solely 
due to differences in the Wannier functions shape. 
Indeed, their real-space representations show marginal differences 
(see SM Fig.~\ref{SM-fig:mlwf_nio2}).
Therefore, the Wannier functions' shape cannot be responsible 
for the different $U$ values found for C1 and C2.
Instead, the origin must be in the electronic configurations 
adopted by Ni($d$) upon the formation of the oxide chain.
To confirm this interpretation, we have analyzed 
the bare and screened Coulomb matrix elements obtained in cRPA calculations 
with different choices of the constrained subspace. 
The $\uparrow \uparrow$ spin channel values of the elements 
needed in the determination of $U$, i.e.\ $U_{mn,mn}$, 
are summarized graphically in Fig.~\ref{fig:check_nio2}
with the contracted index notation $U_{mn}$.
Using as baseline wavefunctions those of the C1 and C2 states 
at the previously converged $U$ values [panels (a,d)], 
the new $U_{mn}$ have been calculated with subspaces that 
include also the O($s$) [panels (b,e)] and O($p$) orbitals [panels (c,f)]. 
This means that in panels (a,d) the screening is (predominantly) due to O($sp$), 
in panels (b,e) to O($p$) and in panels (c,f) only to O($s$).
As a reference, the bare Coulomb matrix elements are shown in panel (g).
The general trend is that the formation of Ni-O bonds largely screens the 
atomic Coulomb interaction at the Ni($d$) orbital, reducing it 
from values $\sim 25$ to $\sim 6.5$ eV in C1 and to $\sim 2.5$\,eV in C2.
The comparison of data columns (a) {\it vs.} (b) and (d) {\it vs.} (e)
in Fig.~\ref{fig:check_nio2} shows that, for both configurations, 
the screening contribution of the O($p$) electrons apparently suffices to
reproduce the complete screening. This does not mean that there is 
no significant O($s$) contribution, though. In panels (c,f) 
we see that the $U$ value for $d$ electrons [$\hat U(d,d)$] is screened by O($s$) electrons 
from 25 to 15\,eV and to 10\,eV in C1 and C2, respectively.
We draw two conclusions from this result: 
(i) the Coulomb screening of $s$ and $p$ channels on Ni($d$) is not additive and
(ii) the  $p$ screening is felt differently in C1 and C2 
because each multiplet binds differently to the neighboring O atoms.

The DOS at the Fermi level $\rho(\epsilon_F)$ is usually taken as an indicator 
of the correlation strength. As a rule of thumb, 
the higher the DOS, the stronger the screening.
Unlike in C2, in the C1 configuration the Fermi level lies 
in a spin-majority bandgap, i.e., C1 is a half-metal, 
which is consistent with a larger effective $U$ value 
(note also that the ZSA theory allows for non-insulating 
behavior despite $U$ being large \cite{bib:zaanen85}). 
This qualitative DOS difference
is also manifested in the spin dependence of the interaction screening:
the $U$ parameters of metallic C2 show an almost negligible 
spin dependence, in contrast to the other half-metallic and insulating chains 
(see the comparison in the SM Table~\ref{SM-tab:Uspinchannels}). 
As shown in Fig.~\ref{fig:bands_nio2}, 
the main contribution to $\rho(\epsilon_F)$ has O($p$) and   
partial $d_{yz}$ character, which indicates conduction along a $\pi$ band.
In C1 and C2 the $d_{xy}$ band is localized in energy and split across the Fermi level 
in a similar manner.
The main difference between the two configurations is in the $d_{z^2,x^2-y^2}$ band
\footnote{
Due to the chain symmetry, bands stemming from $d_z^2$ and $d_{x^2-y^2}$ 
atomic orbitals hybridize. In fact, as shown in Fig.~\ref{fig:bands_nio2}, 
anticrossing features appear for bands with weight in the MLWFs 
corresponding to $d_z^2$ and $d_{x^2-y^2}$, while not on bands with 
$d_{xz}$, $d_{yz}$ or $d_{xy}$ character, as the latter belong to different 
symmetry representations.
},
\newcounter{foot:sym}
\setcounter{foot:sym}{\value{footnote}}  
which is partially filled and strongly hybridized with the O($p$) in C1,
while it is fully filled and localized in C2 
(as it is partially filled in the C1 spin minority channel, 
it results in the different net spin moments of C1 and C2 configurations,
as shown in Table~\ref{tab:occup_nio2}).
This relates to the aforementioned conclusion point (ii) 
in the interpretation of the screening channels. 
Indeed, the bindings to the ligand differ qualitatively: 
the sharp peaks in the DOS around $E_F-1$\,eV form a $d-d$ gap for C2, 
in contrast to the $d-p$ gap of C1 (see Fig.~\ref{fig:bands_nio2} and SM Fig.~\ref{SM-fig:pdos_nio2}), 
which suggests that the C2 configuration is prone to undergo a Mott transition \cite{bib:zaanen85}.

The existence of high-$U$ (C1) and low-$U$ (C2) regimes can 
be rationalized by the estimates of the correlation energy at the Ni($d$) orbitals.
These can be expressed as $E_U = U \sum_i n^{\uparrow}_in^{\downarrow}_i$,
where the $n^{\sigma}_i$ are taken from 
the individual $d$-orbital occupancies of Table~\ref{tab:occup_nio2}.
Owing to the aforementioned symmetry argument \cite{Note\arabic{foot:sym}}, 
we have considered 
$d_{z^2,x^2-y^2}$ as an individual orbital with doubled maximum occupancy
in the $n^{\sigma}_i$ integrals shown in Table~\ref{tab:Unn}.
This simple argument provides the correct tendency obtained by cRPA 
for C1 and C2: the larger occupancy factor of C2 as compared to C1
suggests that the C2 is compatible with a smaller $U$ value than C1. 
The $E_U$ values themselves, namely 25.8\,eV for C1 and 12.2\,eV for C2, 
are very different. 
These energy estimates cannot be used to tell which configuration is more 
stable, as DFT+$U$ total energies obtained with different $U$ values 
cannot be directly compared
\footnote{Since in DFT+$U$ the Hubbard terms are applied to a subset of 
Kohn-Sham eigenstates, which have no physical meaning on their own,
only total energies obtained in same-$U$ calculations can be compared. 
For the particular values $U=6$ and $J=1$\,eV for Ni($d$), 
we find that C2 is metastable with an energy difference of 0.33\,eV 
with respect to C1.}.

\begin{table}[tb!]
\caption{\label{tab:Unn} $n^{\uparrow}n^{\downarrow}$ factors, correlation energy estimates $E_U$ and
and spin moments of the transition metal atom in free-standing and supported chains.}
\begin{ruledtabular}
 \begin{tabular}{c  c  c  c}
   &  $n^{\uparrow}n^{\downarrow}$ &  $E_U$ & $\mu_{X}$ $(\mu_{\text{B}})$\\
\hline
NiO$_2$ C1 & 3.98 & 26.2 & 1.23 \\
NiO$_2$ C2 & 5.09 & 12.3 & 0.55  \\
NiO$_2$/Ir & 5.08 &  8.7 & 0.52 \\ \hline
MnO$_2$    & 0.40 &  1.3 & 3.49 \\
MnO$_2$/Ir & 0.61 &  2.3 & 3.67 \\ \hline
CoO$_2$    & 2.11 & 12.1 & 2.21 \\
CoO$_2$/Ir & 2.74 &  6.5 & 2.02 \\ \hline
FeO$_2$    & 0.80 &  6.1 & 3.65 \\
FeO$_2$/Ir & 1.93 &  2.7 & 2.87
 \end{tabular}
\end{ruledtabular}
\end{table}

\begin{figure}[tb!]
\centerline{ \includegraphics[width=1\columnwidth]{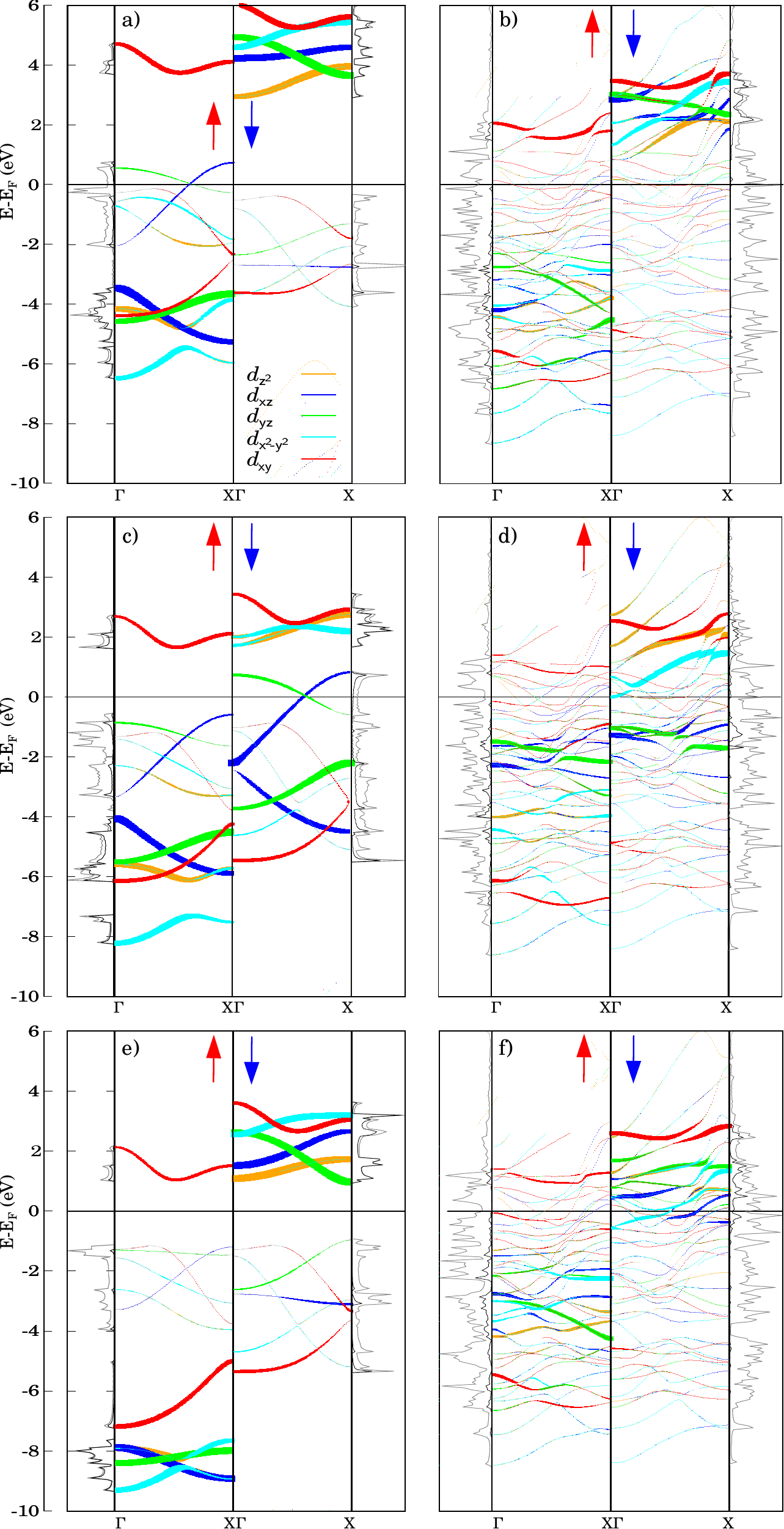}}
\caption{
The same information as in Fig.~\ref{fig:bands_nio2} is provided
here for free-standing planar (left panels) and supported
(right panels) XO$_2$ chains with X = Mn (a,b), Co (c,d) and Fe (e,f).}
\label{fig:bands_others}
\end{figure}

The results of Table~\ref{tab:UJ} follow the trend laid down by the 
DOS at the Fermi level, shown in Fig.~\ref{fig:bands_others} and SM Fig.~\ref{SM-fig:pdos_others} for the 
other studied isolated chains: 
FeO$_2$ is insulating (its band structure shows the features of a conventional CT insulator) 
and, thus, screening is the weakest ($U=7.67$\,eV) among the different studied systems, 
while it is somewhat stronger
in the half-metallic systems MnO$_2$, CoO$_2$ and NiO$_2$-C1 (with $U$ values close to 6\,eV), 
and even stronger in metallic NiO$_2$-C2 ($U=2.41$\,eV). 
In Ref.~\cite{bib:solovyev94} an expression of the screened $U$ is given as 
the derivative of the Kohn-Sham potential with respect to the number $n_d$ 
of electrons in the $d$ orbital, which makes clear the explicit dependence 
on the relaxation of the $d$ bands themselves for a given occupancy. 
When this expression is applied to the case of TM impurities of the same valence state embedded 
in an alkali metal (Rb), screening increases linearly with $n_d$, 
and it is stronger for monovalent than for divalent impurities. 
In the four XO$_2$ isolated chains
this scaling does not apply, because of the more complex chemical environment, 
involving directional bonds and orbital-specific band dispersion. 
Indeed, while the $d$-shell occupancies are similar in the C1 and C2 
configurations ($n_d(\mathrm{C1})=7.71$ and $n_d(\mathrm{C2})=7.92$), the $U$ 
parameters are clearly different (Table~\ref{tab:UJ}),  
due to the pronounced screening contribution from $s$ and $p$ electrons.

\begin{figure}[tb!]
\centerline{\includegraphics[width=0.8\columnwidth]{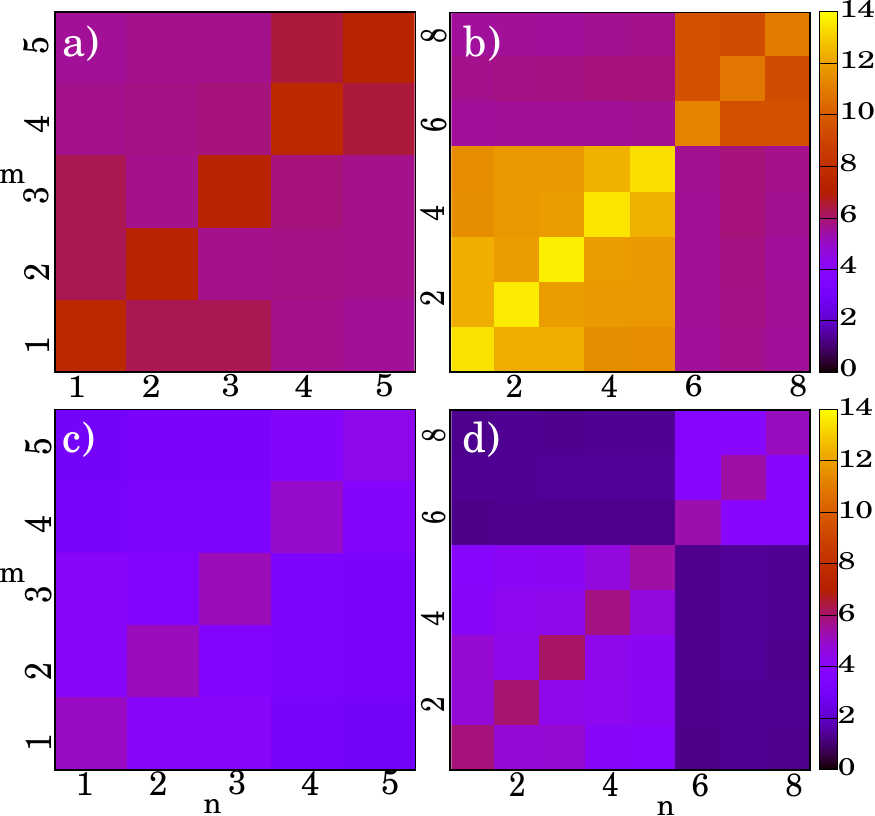}}
\caption{
Effective $U_{mn}$ Coulomb matrix elements (in eV) in the isolated MnO$_2$ chain
including (a) only Mn($d$) electrons and (b) Mn($d$) and O($p$) electrons
in the correlated subspace.
Panels (c,d): same quantities for the supported MnO$_2$ chain 
(Note, O($s$) electrons, which belong to the correlated space in panel (d), are not shown).}
\label{fig:check_mno2}
\end{figure}

It has been argued in Ref.~\cite{bib:seth17} that considering a correlated subspace 
of $d$-states in a system as the present chains, 
where the TM $d$ orbitals strongly hybridize with the O($p$) ligands, 
often leads to an underestimation of the $U$ values. 
In these chains, a bonding-antibonding $d_{xy}-p_{x,y}$ pair is formed, which 
appears as a one-dimensional dispersive occupied band
of width $\sim 2$\,eV (see red lines in Figs.~\ref{fig:bands_nio2} and \ref{fig:bands_others}). 
It is also manifested in the MLWF corresponding to the $d_{xy}$ orbital,
which acquires a nodal feature between the atoms 
(see the FeO$_2$ and NiO$_2$ cases in the SM 
Figs.~\ref{SM-fig:mlwf_feo2} and \ref{SM-fig:mlwf_nio2}, respectively).
In principle, by having this localized wavefunction in the correlated space for cRPA,
the splitting between the X($d$) space and the rest 
retains a partial contribution of the O($p$) ligands in the 
correlated subspace \cite{bib:miyake09}.
Alternatively, when treating $p$ and $d$ electrons as correlated, like in 
Fig.~\ref{fig:check_nio2}(c,f), the off-diagonal blocks $\hat U(d,p)$ account 
for intershell interactions that renormalize the intrashell interactions $\hat U(d,d)$ and $\hat U(p,p)$. 
This is the so-called "shell-folding" approach \cite{bib:seth17}, whereby, 
if the total occupation of the $d$ and $p$ subspaces remains invariant under 
changes in $U$, the renormalization is written simply as $\hat{\tilde U}(d,d) = \hat U(d,d) - \hat U(d,p)$.
This expression corrects the contribution of itinerant electrons to screening. 
With this approach we obtain $\tilde U > U$ (see Table~\ref{tab:UJ} 
and further details in SM Table~\ref{SM-tab:shellfolding} and 
Figs.~\ref{SM-fig:check_others}). 
In the case of MnO$_2$ the difference is only 0.36\,eV, 
which means that the ligand is almost fully disentangled from the correlated $d$ subspace 
(Fig.~\ref{fig:check_mno2}(a,b) shows the $U_{mn}$ matrix elements),
while for NiO$_2$-C2 the difference is as large as 4.62\,eV. 
Moreover, we find an average $U(d,p)=3.29$\,eV in NiO$_2$-C2, while 
it is $\sim 6$\,eV for the other four cases. We can interpret this as the intershell interactions
playing a distinct role in the screening in the C2 configuration. 
 

\subsection{XO$_2$/Ir(100)-$1\times 3$}
\label{sec:irchains}

For oxide films on a metal, the Coulomb interaction can be modelled as the 
difference between the ionization and affinity energies, each corrected by the image 
potential energy $E_{im}$ created by the charged X($d$) shell \cite{bib:duffy83,bib:altieri99}, 
$U = E(d^{n-1}) + E(d^{n+1}) - 2 E(d^n) - 2E_{im}$,
where each total energy term is calculated for a fixed population of the $d$ orbital.
When the XO$_2$ chains are adsorbed on the Ir(100) substrate 
there is, indeed, a reduction in the screened $U$ values  
with respect to the planar free-standing chains in the range $0.7-6.3$\,eV, being
strongest for  FeO$_2$. These results, obtained after 3-4 convergence cycles 
with cRPA, are shown in Table~\ref{tab:UJ}.
Figs.~\ref{fig:bands_nio2} and \ref{fig:bands_others} show that 
$\rho(\epsilon_F)$ is mainly formed by Ir bands, which is a signature of 
the screening effect of the substrate.
Hybridization between chain orbitals and Ir states, 
clearly visible in the band structures, results in a non-uniform 
charge redistribution at the interface. The reduction of the 
intraatomic Coulomb interaction in X($d$) will depend on the 
interfacial electronic structure details, 
which can be captured by {\it ab initio} methods, 
but not by an image charge model potential
\footnote{The image potential tail behaves as $-q/4(z-z_\mathrm{X})$, 
where $q$ is the X($d$) charge, $z_\mathrm{X}$ is the X atom
adsorption height and $z$ the vertical distance from the surface. 
However, the X atom is too close to the surface for that law to be applicable, 
since it is partially inserted in the trough of the Ir missing row 
($z_\mathrm{X}$ values lie in the 0.7-1.25\,{\AA} range \cite{bib:ferstl16}).}.
%
%
Charge transfer upon adsorption modifies the X($d$) orbital occupancies, 
eventually forcing it to adopt a very different multiplet configuration 
from that of the isolated case. As we have seen in the previous section, 
this can have a dramatic effect on the screening contribution of the $p$ ligands.
In brief, the overall screening observed upon adsorption may have a 
non-negligible indirect contribution from a multiplet change.
The importance of this additional mechanism depends on the chain composition.
We note in passing  that the non-coplanar atomic geometry adopted on the Ir(100)
substrate alters the bandstructure of the chains, but this does not 
suffice to produce significant variations in the $U,J$ values with respect to
the planar geometry (we have checked that the 
$U$ values obtained by cRPA for buckled MnO$_2$ free-standing chains,  
where the atomic positions are set as in the adsorbed geometry,
are not significantly affected by this distortion, although it does 
change the bands dispersion, as shown in the SM Fig.~\ref{SM-fig:bands_mno2_buckled}).

We start by describing the case of Mn and Co oxide chains, as they share 
some common features. The substrate modifies the Mn and Co $d$-band  
energies by hybridization, causing in turn the filling of the partially 
occupied $d_{xz,yz \downarrow}$ bands and a subtle accumulation of 
$d_{xy \uparrow}$ states at the Fermi level (see Fig.~\ref{fig:bands_others}).
In the Co case, there is also an additional partial filling of 
$d_{x^2-y^2 \downarrow}$ states (see SM Table~\ref{SM-tab:occup_others}). 
The $n^{\uparrow}n^{\downarrow}$ factors 
for the Mn($d$) and Co($d$) orbitals undergo a weak increase,
compatible with a drop in the $U$ value, as shown in Table~\ref{tab:Unn}. 
%
%
Although the X($d$) electronic configuration 
changes during adsorption are subtle, the interplay of 
the above-mentioned indirect mechanism, driven by the concomitant 
changes in the ligand effects, cannot be ruled out 
until the many-body interactions are examined.
Here, an analysis of the selective inclusion of screening 
channels and shell-folding can be helpful.
Fig.~\ref{fig:check_mno2} shows that, in the MnO$_2$/Ir(100) case,
the screening by Ir atoms only (panel d) yields an intra-Mn($d$)
average interaction $U^{\mathrm{Ir}}(d,d) = 4.66$\,eV.
This value is then renormalized by interaction with O($p$) orbitals
to $\tilde U^{\mathrm{Ir}} = 3.29$\,eV,
which is close to the value $U=3.78$\,eV obtained by the regular method (panel c). 
This implies that for the adsorbed MnO$_2$, as in the isolated case, the
ligand orbitals are well disentangled 
and the screening by the Ir substrate causes the interaction to 
be further reduced by $\simeq 1$\,eV. 
This estimation is obtained as $U^{\mathrm{Ir}}(d,d) - U = 0.88$\,eV 
or $U^{\mathrm{Ir}}(d,d) - \tilde U^{\mathrm{Ir}} = 1.37$\,eV.


%

Next, we consider the Ni oxide chains.
In the adsorbed state, the Ni($d$) configuration resembles better 
that of C2 than C1 (see Tables~\ref{tab:occup_nio2} and ~\ref{tab:Unn}, and Fig.~\ref{fig:check_nio2}). 
Since the C1 configuration is not found on the surface (not
even as a metastable state), we interpret this fact as a consequence 
of the multiplet effect playing a dominant
role in the screening of Coulomb interactions in NiO$_2$/Ir(100), 
leading to the low value of $U=1.71$\, eV (calculated for the $\uparrow\uparrow$ channel).
Indeed, to probe the robustness of the result on Ir,
we have initialized the cRPA iterations with high and low $U_0$ values and
Ni($d$) frozen C1 and C2 configurations. In all cases, the calculation converges
towards the $U$ values and occupancies shown in Tables~\ref{tab:UJ} and \ref{tab:occup_nio2}.
The common feature with Co and Mn is that the filling of $d_{xz,yz}$ bands 
in C2 is completed by charge transfer from the substrate, as shown by 
visual inspection of the band structure (Fig.~\ref{fig:bands_nio2}). 
Ni($d$) states remain overall as narrow bands around the Fermi level.
In particular, the hybrid $d_{z^2,x^2-y^2}$ states, which form a narrow band 
at $E_F-1$\,eV in the isolated NiO$_2$ chain, become partially occupied 
by a relatively weak hybridization with Ir states 
and, therefore, screening is expected to be enhanced. 
All in all, the multiplet features that lead to the low $U=2.41$\,eV value in
isolated NiO$_2$-C2 are present also in the NiO$_2$/Ir(100) adsorbed case. 
The further reduction to $U=1.71$\,eV is compatible with the 
already mentioned hybridization mechanism, 
consisting of a reduction of $\simeq 0.9$\,eV, as estimated from the MnO$_2$/Ir(100) case. 
The same calculation for NiO$_2$/Ir(100), shown in Table~\ref{SM-tab:shellfolding}, 
yields a smaller reduction of $U^{\mathrm{Ir}}(d,d) - U = 0.28$\,eV, 
probably due to the stronger entanglement with O($p$) ligands detected at NiO$_2$-C2.

Finally, we address the FeO$_2$ case.
A very efficient overall screening, with a dramatic reduction of 
the $U$ value from 7.67 to 1.38\,eV, is found in adsorbed FeO$_2$/Ir(100). 
The Fe($d$) multiplet is changed by interaction with 
the Ir substrate, changing from a $S=2$ to a $S=3/2$ state 
(see Table~\ref{tab:UJ}). This is also reflected in the increase by 
one unit in the $n^\uparrow n^\downarrow$ factor.
The FeO$_2$ bandstructure, which is that of a CT insulator when the 
chain is isolated, becomes conducting upon hybridization with Ir, 
as the Fe($d$) spin-minority band bottom edge is pinned to the Fermi level. 
We attribute the strong $U$ reduction of the Fe($d$) shell 
to this insulating-to-metallic transition.
The pinning of the Fe($d_{\downarrow}$) band at the Fermi level
persists when high $U$ values are used in the band structure calculation.

Surprisingly, the pinning also persists when the chain is artificially lifted from the substrate.
To understand the evolution of the substrate screening length, we have carried out cRPA
for structures with intermediate height $z_\mathrm{Fe}=2.5$\,{\AA} 
and two initial $U_0$ values to prevent stagnation at metastable states.
Starting from both $U_0^{f}=7.67$\,eV (the converged free-standing value)
and $U_0^{s}=1.38$\,eV (the converged supported value), 
the value is stabilized at $U_2^{f,s} \simeq 3.5$\,eV after two iterations 
and a residual peak of chain states is still visible at the Fermi level.
A height as large as $z_\mathrm{Fe}=4$\,{\AA} is needed for 
these residual states to vanish, as shown in SM Fig.~\ref{SM-fig:pdos_feo2_z}(d)
(the PDOS corresponding to some of those cases is shown in SM Fig.~\ref{SM-fig:pdos_feo2_z}).
However, at this distance, the $U$ parameter does not 
stabilize at the value of the free-standing chain $U_0^{f}$, 
but reduces to $U_1^{f}=6.13$\,eV after one iteration. 
Starting from the adsorbed value $U_0^{s}$, it results in $U_1^{s}=3.81$\,eV. 
Therefore, despite the hybridization of the chain states with the surface 
is negligible at $z_\mathrm{Fe}=4$\,{\AA}, a non-negligible screening persists. 
This behavior is not necessarily unphysical.
Nevertheless, in order to accurately describe the screening for far-lying 
chains, a different exchange and correlation ($xc$) functional would be needed. 
Note, the GGA, being a semilocal $xc$ functional, works by error cancellation, 
providing a poorer description when low electron densities are involved in the interactions,
failing to provide an accurate asymptotic $1/z$ behavior.


\section{Conclusions}
\label{sec:conclusions}

In summary, we have performed a cRPA investigation of magnetic one-dimensional 
transition metal oxide XO$_2$ chains deposited on a Ir(100) surface aimed at 
understanding the screening of intraorbital Coulomb interactions in the X($d$) shell  
under  the combined effect of low dimensionality, the ligand field and a 
neighboring metal.

Calculations for the isolated XO$_2$ chains show a strong dependence 
of the Hubbard $U$ parameter on the X species, ranging 
from 2.4 to 7.7\,eV. 
In each case, the $U$ reflects the insulating or 
(half)metallic character of the chain. Importantly, we find low-$U$ and 
high-$U$ regimes in the case of NiO$_2$ associated to $d-d$ (Mott-Hubbard) 
or $p-d$ (charge transfer) gap types, respectively.
The gap type is determined by the Ni($d$) electronic configuration or 
multiplet. In the particular case of NiO$_2$, multiplets with even 
the same orbital filling and same spin state 
(as it is the case of the C1 and C2 multiplets found in 
the present work) lead to different regimes.
Due to the interaction with the O($p$) ligands, 
the inclusion of O($p$) electrons in the cRPA correlated subspace 
results in a different renormalization 
and higher values of the $U$ parameter
for each transition-metal species. The increase is
smallest for MnO$_2$ and largest for the NiO$_2$ in the Mott-Hubbard-like multiplet.

Since the ligand field is weak for MnO$_2$, this case allows 
us to establish that the $U$ reduction by interaction with the metallic 
substrate is $\simeq 0.9$\,eV and a much larger reduction comes
from the change of the occupation of the $d$-states. 
In general, however, the contributions of substrate and 
ligand cannot be uncoupled. Adsorption drives the formation of interfacial states
and charge transfer to the chain, which in turn can 
undergo an insulator-to-metal transition
(such is the FeO$_2$ case, where the value of $U$ 
is reduced by almost 6\,eV) or have its multiplet configuration altered.
The latter is the scenario for NiO$_2$/Ir(100), where 
a low value $U=1.71$\,eV is obtained in part because 
the substrate adopts the Mott-Hubbard-like multiplet. 
Incidentally, the fine details of the band dispersion  
play a lesser role in the determination of the $U$ parameter.

All in all, Coulomb interactions in a low-dimensional
oxide by a neighboring metal cannot be described in simple terms by 
a charge screening model. Instead, the fine details of the hybrid
oxide-metal electronic structure must be considered, as they 
also affect the renormalization of the interactions due to the 
O($p$) ligands.

\begin{acknowledgments}
Projects PID2019-103910GB-I00
funded by MCIN/AEI/10.13039/501100011033/;
GIU18/138 by Universidad del Pa\'{\i}s Vasco UPV/EHU;
IT-1246-19 and IT-1260-19 by Gobierno Vasco.
Computational resources were partially provided by the DIPC computing center.
S.B.\ acknowledges funding from the Deutsche Forschungsgemeinschaft
(DFG) through priority program SPP 2137 “Skyrmionics” (Project BL 444/16) 
and the Collaborative Research Centers SFB 1238 (Project C01).
\end{acknowledgments}

\bibliography{uchains}

\makeatletter\@input{aux4_sm_uchains_v1.tex}\makeatother

\pagebreak


\end{document}



\title{Supplemental Material for:
``Multiplet Effects in the Electronic Correlation
of One-Dimensional Magnetic Transition-Metal Oxides on Metals''}



\newcommand{\QUIMI}[0]{{
Departamento de Pol\'{\i}meros y Materiales Avanzados: 
F\'{\i}sica, Qu\'{\i}mica y Tecnolog\'{\i}a, Facultad de Qu\'{\i}mica UPV/EHU,
Apartado 1072, 20080 Donostia-San Sebasti\'an, Spain}}

\newcommand{\DIPC}[0]{{
Donostia International Physics Center, 
Paseo Manuel de Lardiz\'abal 4, 20018 Donostia-San Sebasti\'an, Spain}}

\newcommand{\CFM}[0]{{
Centro de F\'{\i}sica de Materiales CFM/MPC (CSIC-UPV/EHU), 
Paseo Manuel de Lardiz\'abal 5, 20018 Donostia-San Sebasti\'an, Spain}}

\newcommand{\JUELICH}[0]{{
Peter Gr\"unberg Institut and Institute for Advanced Simulation, 
Forschungszentrum J\"ulich and JARA, 52425 J\"ulich, Germany}}

\newcommand{\ICMM}[0]{{
Instituto de Ciencia de Materiales de Madrid, CSIC,
Cantoblanco, 28049 Madrid, Spain}}

\author{J. Goikoetxea}
\affiliation{\CFM}
\author{C. Friedrich}
\affiliation{\JUELICH}
\author{G. Bihlmayer}
\affiliation{\JUELICH}
\author{S. Bl\"ugel}
\affiliation{\JUELICH}
\author{A. Arnau}
\affiliation{\CFM}
\affiliation{\QUIMI}
\affiliation{\DIPC}
\author{M. Blanco-Rey}
\affiliation{\QUIMI}
\affiliation{\DIPC}



\date{\today}



\maketitle

\clearpage

\begin{table}[tb!]
\caption{\label{tab:struc_free}
\textbf{Relaxed isolated coplanar XO$_2$ structures.}
Equilibrium geometries have been obtained with the code \textsc{VASP},
based in a plane-wave basis set and projector augmented wave potentials \cite{bib:kresse93,bib:kresse99}.
These are DFT+$U$ calculations \cite{bib:anisimov97} with $U=1.5$\,eV and the generalized 
gradient approximation PBE of the exchange and correlation functional \cite{bib:pbe96}.
To model the isolated chains, supercells that leave a 10\,{\AA} vacuum spacing 
between chains are used.
The basis is constructed with a $10 \times 1 \times 1$ sampling of the first 
Brillouin zone and a cut-off of 450\,eV. The convergence thresholds 
are $10^{-5}$\,eV for the total energies and 0.01\,eV{\AA}$^{-1}$ for the forces on
the ions.
The values in the table are the interatomic distances found after 
relaxation of the XO$_2$ chains with the ideal coplanar geometry. }
\begin{ruledtabular}
\begin{tabular}{c  c  c  c  c}
  & MnO$_2$ & FeO$_2$ & CoO$_2$ & NiO$_2$ \\
\hline
$d_{\mathrm{X-X}}$\,({\AA}) & 2.75 & 2.71 & 2.65 & 2.70  \\
$d_{\mathrm{O-O}}$\,({\AA}) & 2.43 & 2.43 & 2.41 & 2.35 \\
$d_{\mathrm{X-O}}$\,({\AA}) & 1.87 & 1.73 & 1.79 & 1.79  \\
\end{tabular}
\end{ruledtabular}
\end{table}

\begin{table}[tb!]
\caption{\label{tab:struc_adsorbed}
\textbf{Relaxed structures of XO$_2$/Ir(100).}
For the supported chains, the known $(3\times 1)$ missing row reconstructed 
structure has been relaxed, using a Ir substrate of five atomic planes, 
where the two bottom Ir planes are kept fixed. 
The unit cell lateral size is fixed by the Ir(100) lattice constant (2.71\,{\AA}), 
and the vacuum spacing between slab replicas is 10\,{\AA}. 
The first Brillouin zone sampling is $10\times 3 \times 1$. 
Calculation details are as in the caption of Table~\ref{tab:struc_free}.
The values in the table are the relaxed interatomic distances $d$, 
the heights of the X and O atoms $z$ measured from the topmost Ir plane,
and the Ir interplanar distances $\Delta_i$ (between the average $z$ 
of atomic plane $i$ and $i+1$), ordered from the topmost plane.
Individual Ir atom displacements, not shown in the table, 
are $\simeq 0.1$\,{\AA} at the topmost plane and $\leq 0.05$\,{\AA} 
at the second and third planes.}
\begin{ruledtabular}
\begin{tabular}{c  c  c  c  c}
  & MnO$_2$ & FeO$_2$ & CoO$_2$ & NiO$_2$ \\
\hline
$d_{\mathrm{O-O}}$\,({\AA}) &  2.62 & 2.63 & 2.64 & 2.65 \\
$d_{\mathrm{X-O}}$\,({\AA}) &  1.90 & 1.90 & 1.96 & 1.85 \\
$z_{\mathrm{O}}$\,({\AA})   &  1.39 & 1.31 & 1.34 & 1.32 \\
$z_{\mathrm{X}}$\,({\AA})   &  1.12 & 1.89 & 0.17 & 1.19 \\
$\Delta_{1}$\,({\AA})       &  1.84 & 1.84 & 1.84 & 1.87 \\
$\Delta_{2}$\,({\AA})       &  1.99 & 1.99 & 1.99 & 1.98 \\
$\Delta_{3}$\,({\AA})       &  1.84 & 1.84 & 1.84 & 1.84 \\
\end{tabular}
\end{ruledtabular}
\end{table}

\begin{table}[tb!]
\caption{\label{tab:flapw_params}
\textbf{FLAPW calculations parameters.}
Parameters used in the DFT+$U$ FLAPW calculations \cite{bib:krakauer79,bib:wimmer81,bib:shick99}
with the code \textsc{Fleur} \cite{bib:fleur}
in order to obtain the wavefunctions used as 
input for the cRPA calculations for the supported chains on Ir(100).
In parenthesis, the parameters corresponding to isolated chain calculations.
$R_{MT}$ are the muffin-tin radii for the local 
basis functions at atoms X (transition metal (TM)), O and Ir, which are 
are expanded up to $l_{max}=8,6$ and 6 angular momentum components, respectively
(non-spherical potential components are expanded up to $l_{max}=6,4$ and 6, respectively).
The cutoffs for the basis functions and potential/density in the interstitial regions are given by 
$E_c^{pot}/E_c^{wvf}$, respectively. The Brillouin zone was sampled by 
a $12\times 3 \times 1$ $k$-point grid.
}
\begin{ruledtabular}
\begin{tabular}{c  c  c  c  c}
  & MnO$_2$ & FeO$_2$ & CoO$_2$ & NiO$_2$ \\
\hline
$R_{MT}^\mathrm{X}$\,(a.u.)  & 2.24(2.16) & 2.28(2.14) & 2.23(2.11) & 2.23(2.11)  \\
$R_{MT}^\mathrm{O}$\,(a.u.)  & 1.27(1.22) & 1.29(1.21) & 1.26(1.19) & 1.26(1.19) \\
$R_{MT}^\mathrm{Ir}$\,(a.u.) & 2.45       & 2.35       & 2.37       & 2.34  \\
$E_c^{wvf}$\,(a.u.$^{-1}$)   & 4.7(4.9)   & 4.5(5.0)   & 4.7(5.0)   & 4.6(4.5)   \\
$E_c^{pot}$\,(a.u.$^{-1}$)   & 14.2(14.7) & 13.8(14.9) & 14.2(14.9) & 14.1(13.0)   \\
\end{tabular}
\end{ruledtabular}
\end{table}

\begin{table}[tb!]
\caption{\label{tab:occup_others}
\textbf{$d$-orbital occupancies and magnetic spin moments of the X atom (X=Mn,Fe,Co).}
Occupations are calculated from integrals of the PDOS curves (see Fig.~\ref{fig:pdos_others}).
The spin moment $\mu_{\mathrm{X}}$ is obtained from the charge inside the X muffin-tin spheres.
}
\begin{ruledtabular}
\begin{tabular}{c  c  c  c  c  c  c  c}
  & Spin & $d_{z^2}$ & $d_{xz}$ & $d_{yz}$ & $d_{x^2-y^2}$ & $d_{xy}$ &  $\mu_{\mathrm{X}}$ ($\mu_B$) \\
   \hline
  \multirow{2}{*}{MnO$_2$} & $\uparrow$    & 0.94  & 0.91 & 0.86 & 0.99 & 0.41  & \multirow{2}{*}{3.49} \\
                           & $\downarrow$  & 0.01  & 0.10 & 0.08 & 0.09 & 0.28 \\
  \hline
 \multirow{2}{*}{MnO$_2$/Ir} & $\uparrow$  & 0.96 & 0.98 & 0.98 & 0.99 & 0.41 & \multirow{2}{*}{3.66}\\
                           & $\downarrow$  & 0.08 & 0.08 & 0.08 & 0.09 & 0.23 \\
  \hline
  \multirow{2}{*}{FeO$_2$} & $\uparrow$    & 0.98  & 0.98 & 1.00 & 1.00 & 0.59  & \multirow{2}{*}{3.67} \\
                           & $\downarrow$  & 0.01  & 0.21 & 0.14 & 0.11 & 0.49 \\
  \hline
 \multirow{2}{*}{FeO$_2$/Ir} & $\uparrow$  & 0.95 & 0.97 & 0.97 & 0.98 & 0.68 & \multirow{2}{*}{2.87}\\
                           & $\downarrow$  & 0.25 & 0.41 & 0.17 & 0.36 & 0.28 \\
  \hline
  \multirow{2}{*}{CoO$_2$} & $\uparrow$    & 0.97  & 1.00 & 1.00 & 0.96 & 0.52  & \multirow{2}{*}{2.21} \\
                           & $\downarrow$  & 0.00  & 0.84 & 0.75 & 0.14 & 0.46 \\
  \hline
 \multirow{2}{*}{CoO$_2$/Ir} & $\uparrow$  & 0.97 & 0.96 & 0.96 & 0.98 & 0.70 & \multirow{2}{*}{2.02}\\
                           & $\downarrow$  & 0.08 & 0.93 & 0.94 & 0.29 & 0.31 \\
\end{tabular}
\end{ruledtabular}
\end{table}

\begin{table}[tb!]
\caption{\label{tab:Uspinchannels}
\textbf{Spin channel effect on the $U$ parameter.}
Values of the $U$ and $J$ parameters calculated from averages over the 
$\uparrow\uparrow$, $\uparrow\downarrow$ and $\downarrow\downarrow$ 
spin channels. Note that the spread in values is only $\approx 0.5$\,eV.
All values in eV. 
The last columns show, for the isolated chain cases, the dimensionless coefficients
$\delta_U = \frac{2(U^{\uparrow\uparrow}-U^{\downarrow\downarrow})}{U^{\uparrow\uparrow}+U^{\downarrow\downarrow}}$ 
and 
$\delta_J = \frac{2(J^{\uparrow\uparrow}-J^{\downarrow\downarrow})}{J^{\uparrow\uparrow}+J^{\downarrow\downarrow}}$,
which accounts for the relative variation in each case as a function of the spin. 
The labels indicate the character of the chain:
HM = half-metallic, M = metallic, or I = insulator. 
Note that the smallest variation of $U$ and $J$ parameters occurs at the metallic NiO$_2$ (C2) chain.}
\begin{ruledtabular}
\begin{tabular}{c  c  c  c |  c c c | c c c}
          & $U^{\uparrow\uparrow}$ & $U^{\uparrow\downarrow}$ &  $U^{\downarrow\downarrow}$ 
          & $J^{\uparrow\uparrow}$ & $J^{\uparrow\downarrow}$ &  $J^{\downarrow\downarrow}$ 
          & $\delta_U$ & $\delta_J$ & \\
\hline
MnO$_2$         &   6.21  & 5.95 & 5.73 & 1.04 & 0.95 & 0.89 & 0.080 & 0.155 & HM \\
MnO$_2$/Ir(100) &   3.78  & 3.53 & 3.33 & 0.98 & 0.88 & 0.83 & & & \\
FeO$_2$         &   7.67  & 7.38 & 7.12 & 1.13 & 1.04 & 0.97 & 0.074 & 0.152 & I \\
FeO$_2$/Ir      &   1.38  &  -   & 1.32 & 0.80 & -    & 0.73 & & & \\
CoO$_2$         &   5.73  & 5.58 & 5.45 & 1.11 & 1.05 & 1.00 & 0.050 & 0.104 & HM \\
CoO$_2$/Ir      &   2.39  &  -   & -    & 0.90 & -    & - & & & \\
NiO$_2$ (C1)    &   6.59  & 6.49 & 6.38 & 1.17 & 1.13 & 1.10 & 0.032 & 0.062 & HM \\
NiO$_2$ (C2)    &   2.41  & 2.40 & 2.39 & 1.01 & 1.00 & 1.00 & 0.008 & 0.001 & M \\
NiO$_2$/Ir      &   1.71  &  -   & 1.70 & 0.87 & -    & 0.86 & & &
\end{tabular}
\end{ruledtabular}
\end{table}

\begin{table}[tb!]
\caption{\label{tab:shellfolding}
\textbf{Shell-folding details.}
Determination of the $\tilde U$ parameter by the shell folding approximation \cite{bib:seth17}
using cRPA calculations where the correlated subspace is formed by the O($p$) and X($d$) orbitals.
The values in the table are averaged over the indicated orbitals ($dd$, $dp$ or $pp$) 
for the spin channel $\uparrow\uparrow$.
The last column shows the difference with respect to the $U$ parameter obtained in the usual 
cRPA calculation, where only the $d$ orbitals form the correlated subspace.
All values in eV. }
\begin{ruledtabular}
\begin{tabular}{c  c  c  c  c  c  c}
            & $U(d,d)$ & $U(d,p)$ & $U(p,p)$ & $\tilde U = U(d,d)-U(d,p)$ & $\tilde U - U$ \\
\hline
MnO$_2$ &        12.24 & 5.67 &  9.96 & 6.57 & 0.36 \\
MnO$_2$/Ir(100) & 4.66 & 1.37 &  4.34 & 3.29 & -0.49  \\
FeO$_2$ &        15.19 & 6.13 & 10.64 & 9.06 & 1.39 \\
CoO$_2$ &        14.84 & 6.22 & 11.08 & 8.62 & 2.89 \\
NiO$_2$ (C1) &   14.61 & 6.15 & 11.26 & 8.45 & 1.86 \\
NiO$_2$ (C2) &   10.31 & 3.28 &  8.77 & 7.03 & 4.62 \\
NiO$_2$/Ir(100) & 1.99 & 0.83 &  1.26 & 1.16 & -0.55
\end{tabular}
\end{ruledtabular}
\end{table}


\clearpage

\begin{figure}[tb!]
\centerline{ \includegraphics[width=0.8\columnwidth]{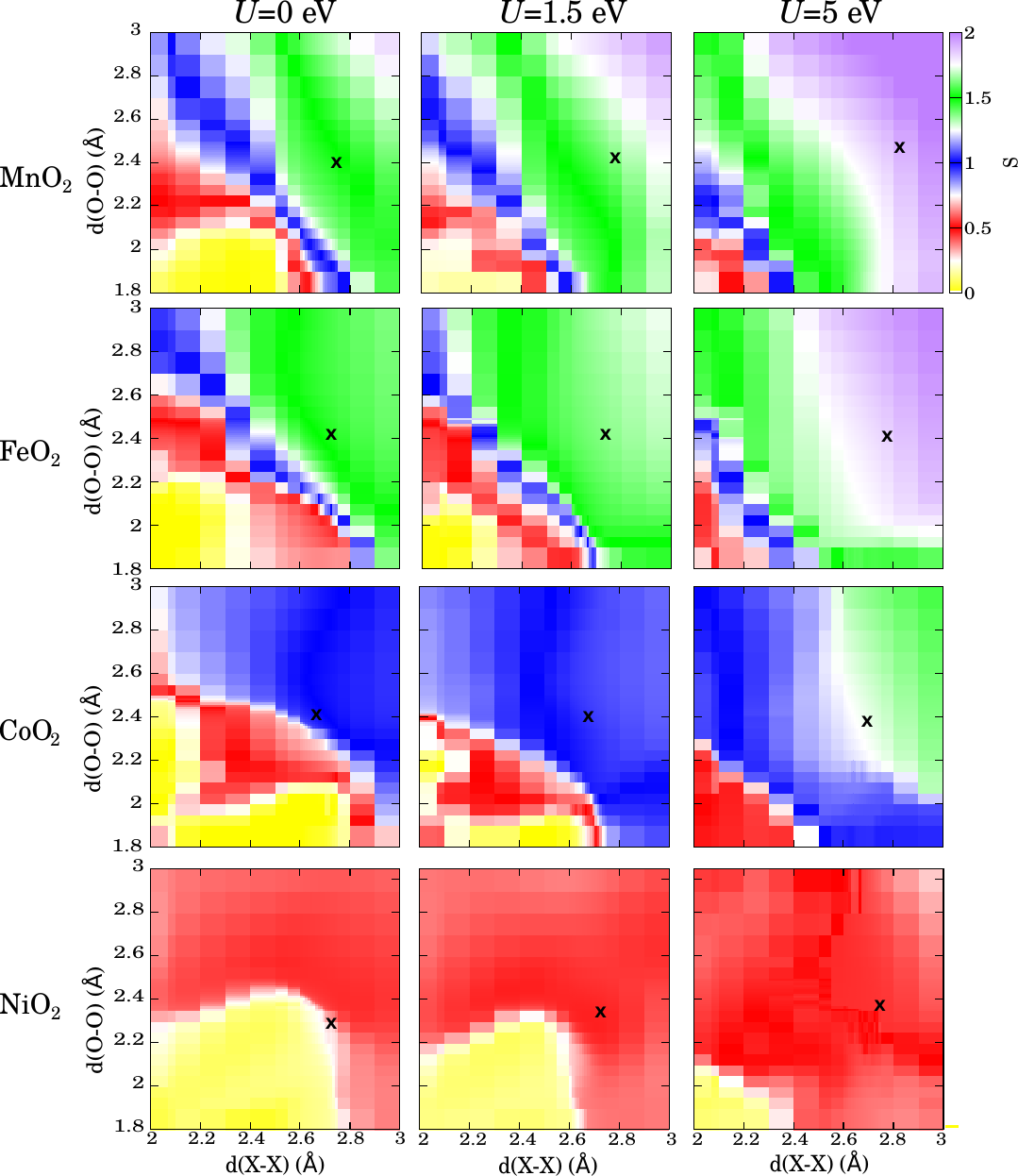}}
\caption{\textbf{Spin states of X in the isolated XO$_2$ chains.}
In these phase diagrams, 
the color code shows the spin values of the X atom, $S$, in isolated coplanar XO$_2$ 
obtained by calculations with VASP (see caption of Table~\ref{tab:struc_free}). 
Each column of panels corresponds to a value of the $U$ parameter 
and each row to a different X atom. In each panel, the X-O and X-X 
interatomic distances are varied and the cross indicates the ground
state geometry for that $U$ value. Note that $U$ has little influence on 
the equilibrium geometry.
As $U$ increases, except for Ni, the spin can change to a higher value. 
In addition, more spin states would become accessible by geometry distortion.
The regions shaded in white correspond to the transitions between spin states. }
\label{fig:spin_phases}
\end{figure}

\begin{figure}[tb!]
\centerline{ \includegraphics[width=0.5\columnwidth]{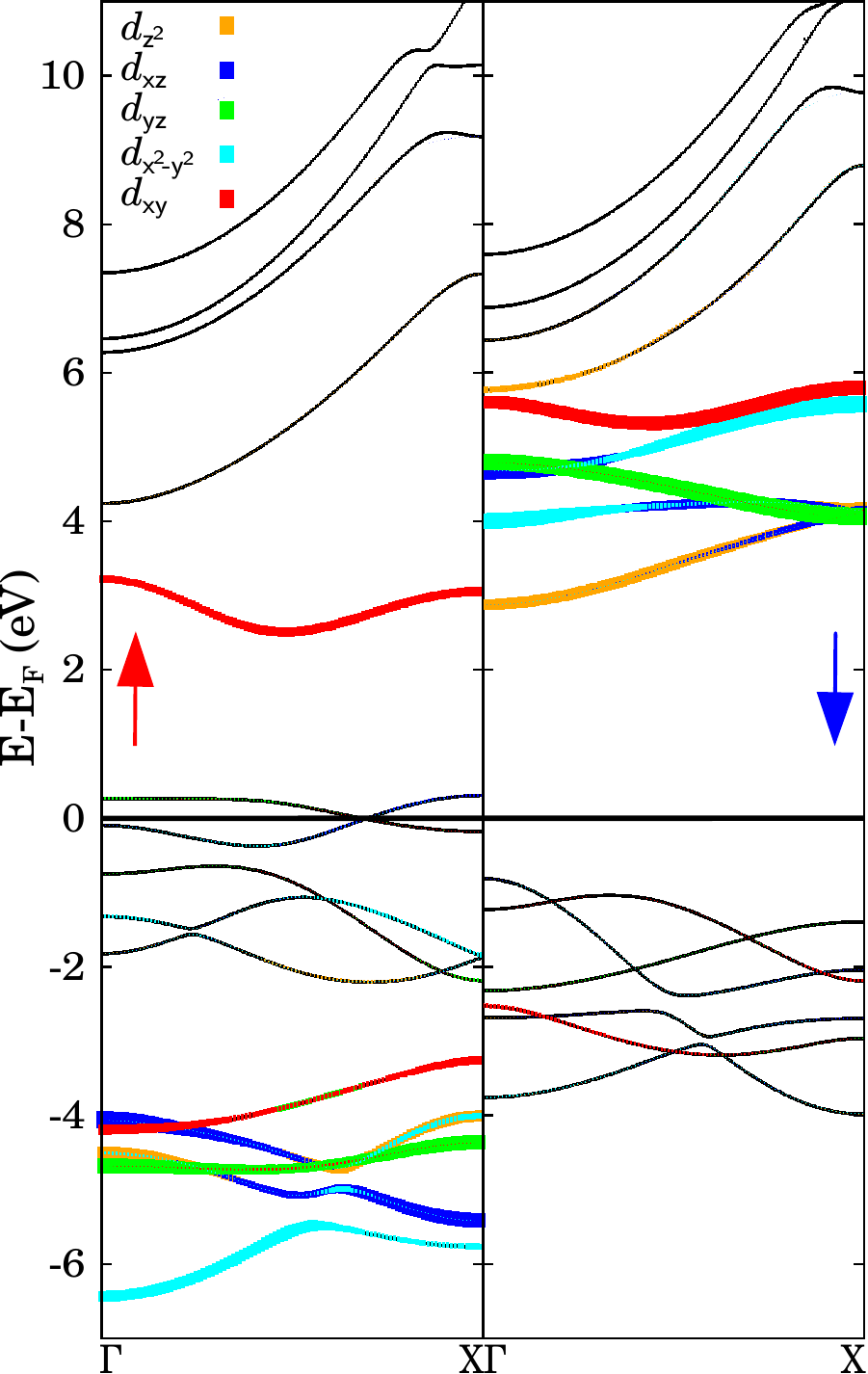}}
\caption{\textbf{Band structure of buckled free-standing MnO$_2$ chains.}
Band structure for free-standing MnO$_2$ in the buckled geometry adopted during adsorption on
the Ir(100) troughs. Left-hand and right-hand panels correspond to 
majority and minority spin polarization, respectively.
The TM's orbital $d$-characters, indicated in the colour code, 
are obtained from the corresponding MLWFs,
with dot sizes representing the projection magnitudes.}
\label{fig:bands_mno2_buckled}
\end{figure}

\begin{figure}[tb!]
\centerline{\includegraphics[width=1\columnwidth]{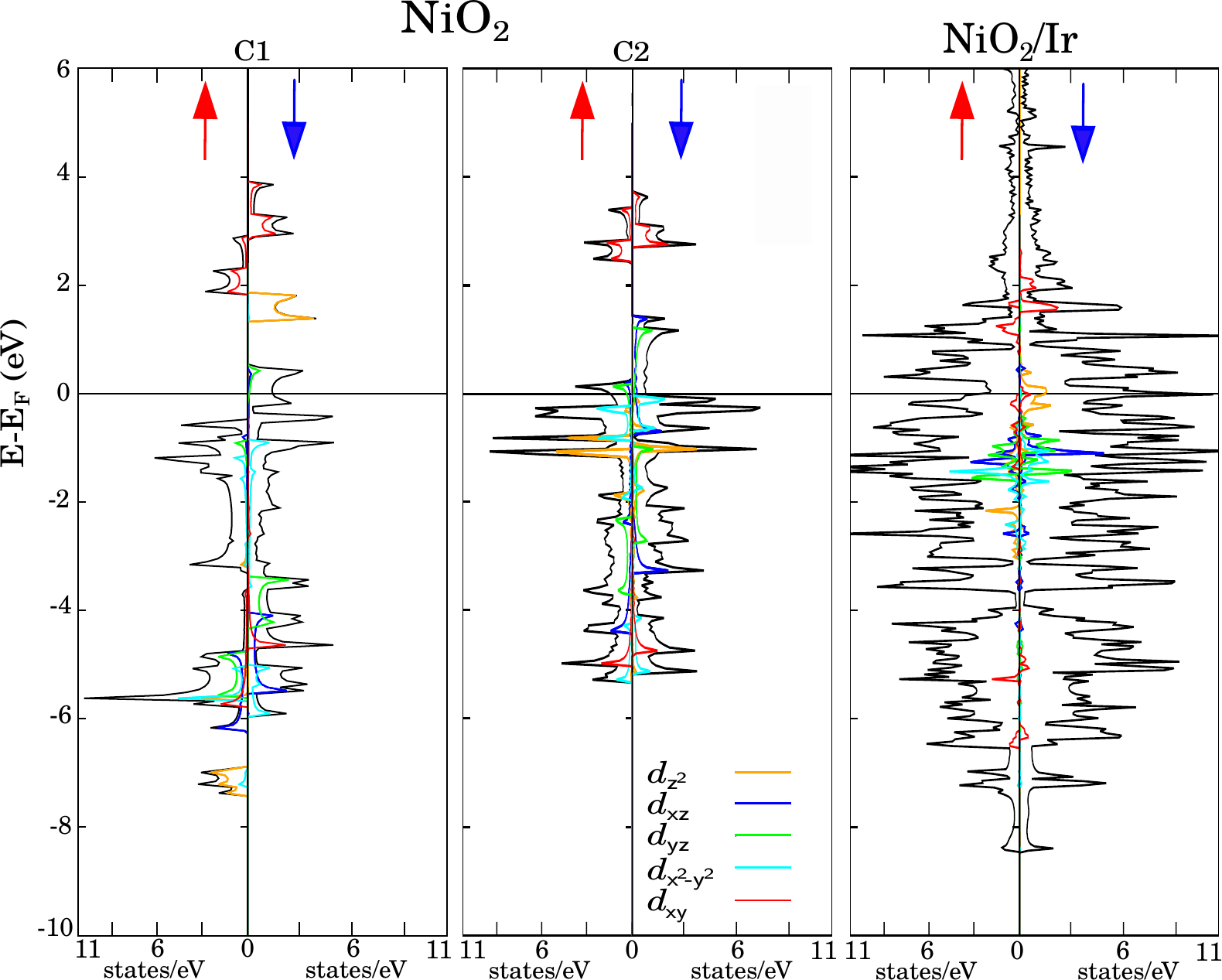}}
\caption{\textbf{Projected densities of states (PDOS) for NiO$_2$ chains.}
In arbitrary units, PDOS of the NiO$_2$ chains resolved in orbital
(colored curves) and spin majority and minority contributions
(panels labelled by red and blue arrows, respectively). 
The black curves account for the total DOS. 
These PDOS correspond to the band structures shown in Fig.~\ref{MP-fig:bands_nio2} of the main text.
The orbital characters are obtained from the corresponding MLWFs.
The free-standing planar NiO$_2$ chain PDOS in the C1 and C2
configurations is shown in the left and middle panels, respectively
and the right-hand panel shows the Ir-supported case,
where the large substrate contribution is contained in the
black curve.}
\label{fig:pdos_nio2}
\end{figure}

\begin{figure}[tb!]
\centerline{ \includegraphics[width=0.8\columnwidth]{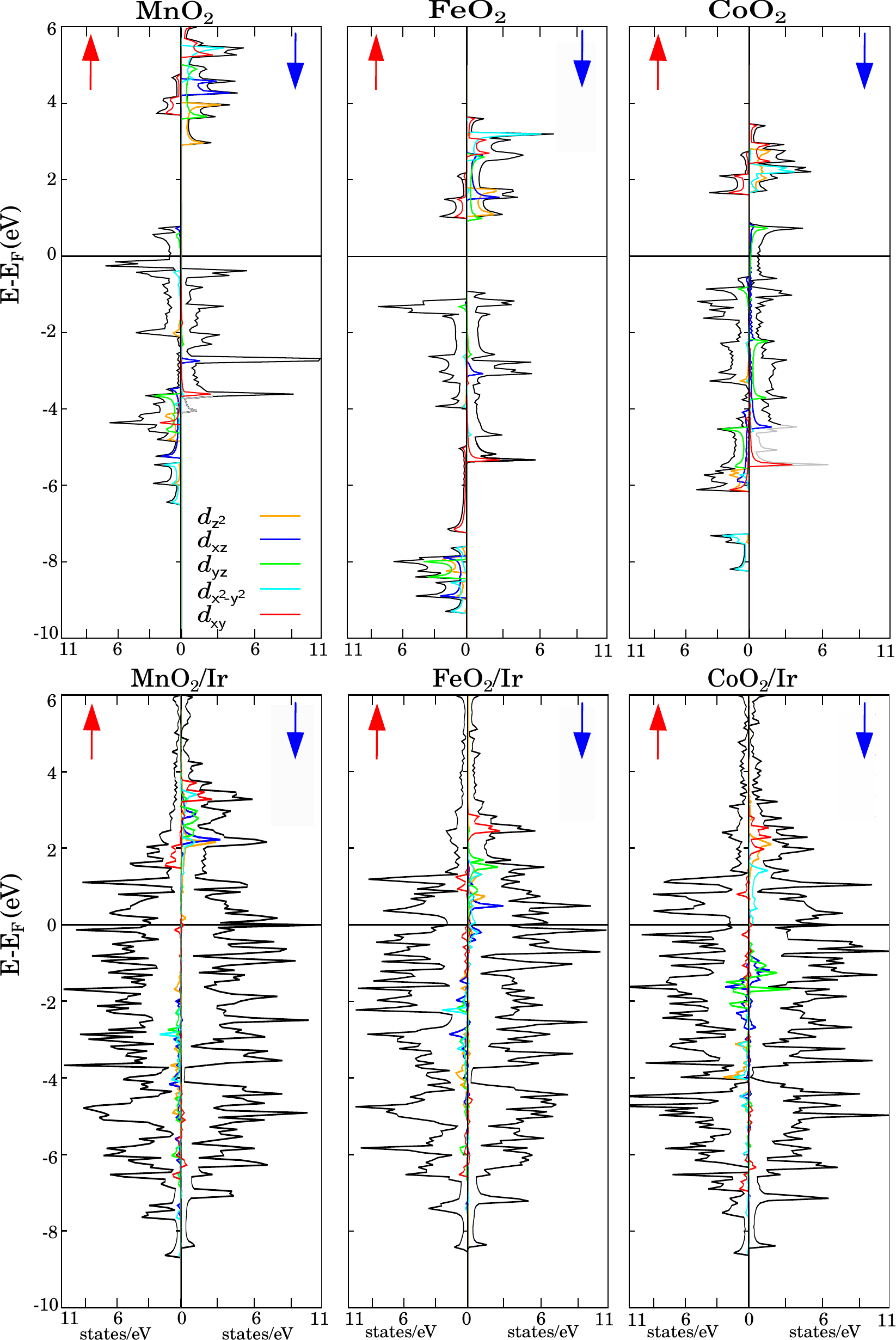}}
\caption{\textbf{PDOS for MnO$_2$, FeO$_2$ and CoO$_2$ chains.}
The same information as in Fig.~\ref{fig:pdos_nio2} 
(corresponding to the band structures shown in Fig.~\ref{MP-fig:bands_others} 
of the main text) is provided here for free-standing planar (top panels) 
and supported (bottom panels) XO$_2$ (X=Mn,Fe,Co) chains.}
\label{fig:pdos_others}
\end{figure}

\begin{figure}[tb!]
\centerline{\includegraphics[width=0.5\columnwidth]{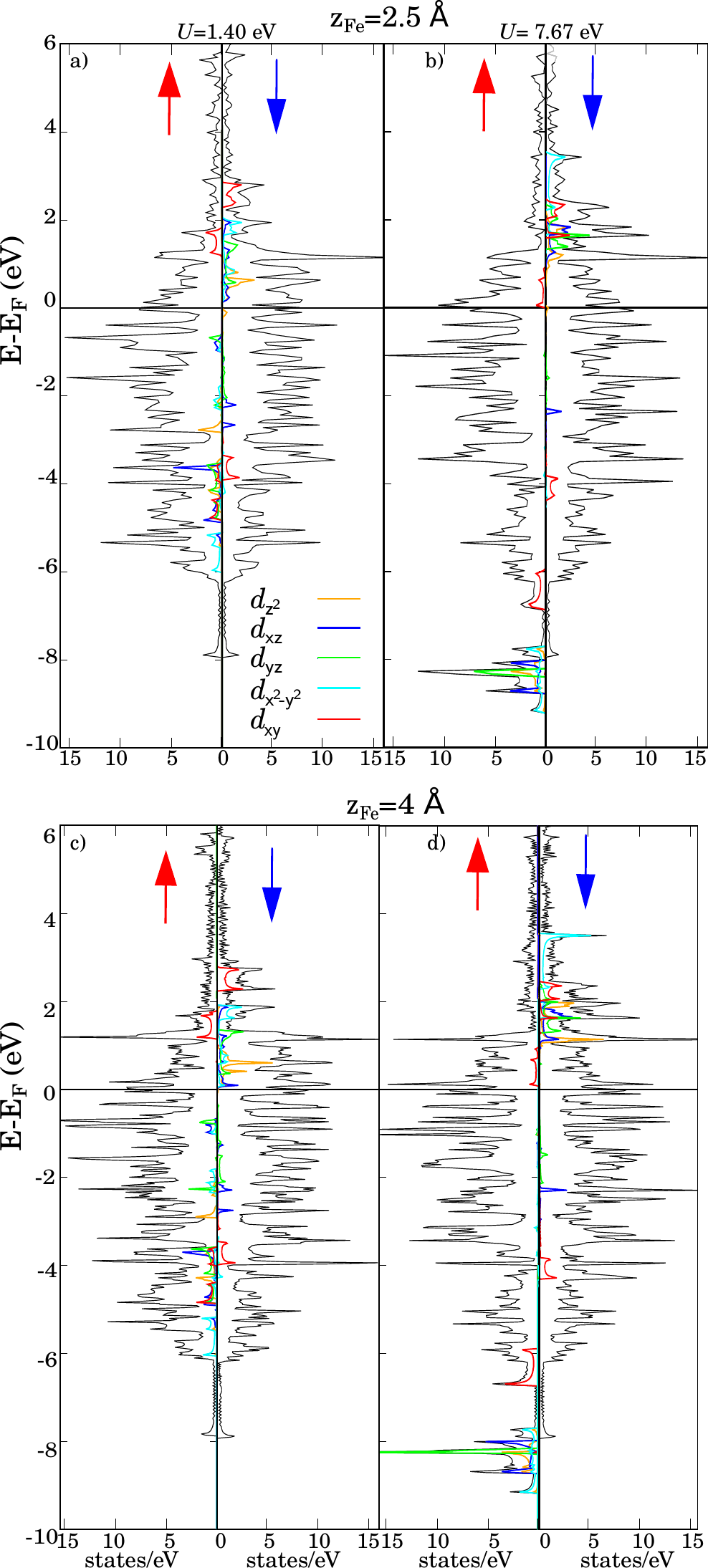}}
\caption{\textbf{Adsorption height dependence of the PDOS for FeO$_2$/Ir(100).}
The colored curves show Fe($d$) orbital contributions to the PDOS 
of FeO$_2$ detached from the Ir(100) substrate at heights 2.5\,{\AA} (a,b) 
and 4\,{\AA} (c,d), for $U=1.40$ and 7.67\,eV, 
which correspond to the limit $U$ values in the adsorbed (a,c) and free standing (b,d) 
configurations, respectively.
The black curve corresponds to the total density of states.}
\label{fig:pdos_feo2_z}
\end{figure}

\begin{figure}[tb!]
\centerline{\includegraphics[width=0.66\columnwidth]{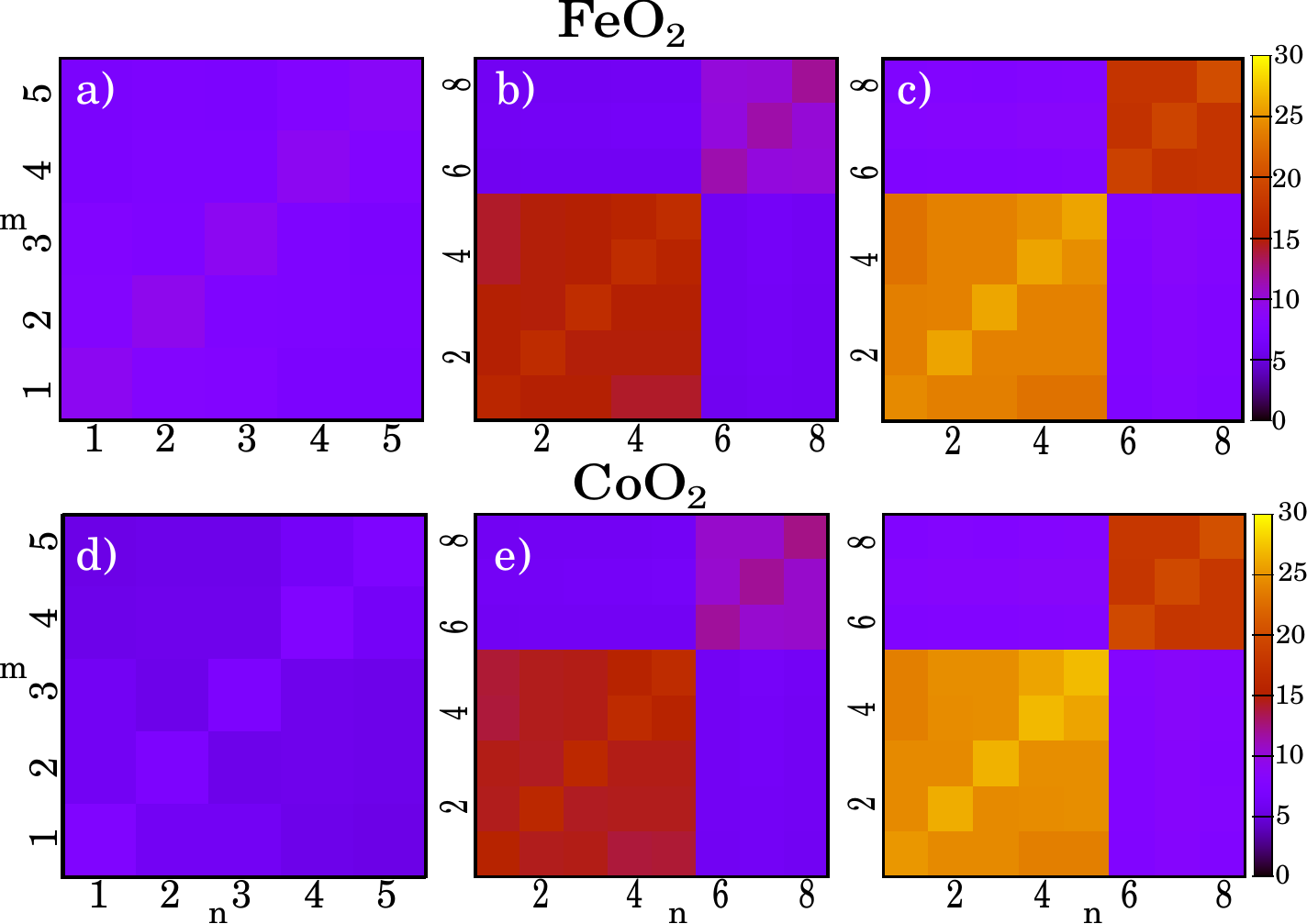}}
\caption{\textbf{Coulomb matrix elements for FeO$_2$ and CoO$_2$ isolated chains.}
(a,d) Effective $U_{mn}$ elements of the X($d$) orbital screened by the rest of electrons.
(b,e) $U_{mn}$ of X($d$) and O($p$) orbitals screened by the rest of electrons ($s$).
(c,f) Bare Coulomb matrix elements $U_{mn}$ for X($d$) and O($p$) in the isolated XO$_2$
chains, with X=Fe (top) and Co(bottom).
Indices are ordered as  $1-d_{z^2}$, $2-d_{xz}$, $3-d_{yz}$, $4-d_{x^2-y^2}$, $5-d_{xy}$,
$6-p_z$, $7-p_x$ and $8-p_y$.}
\label{fig:check_others}
\end{figure}

\begin{figure}[tb!]
\centerline{\includegraphics[width=1\columnwidth]{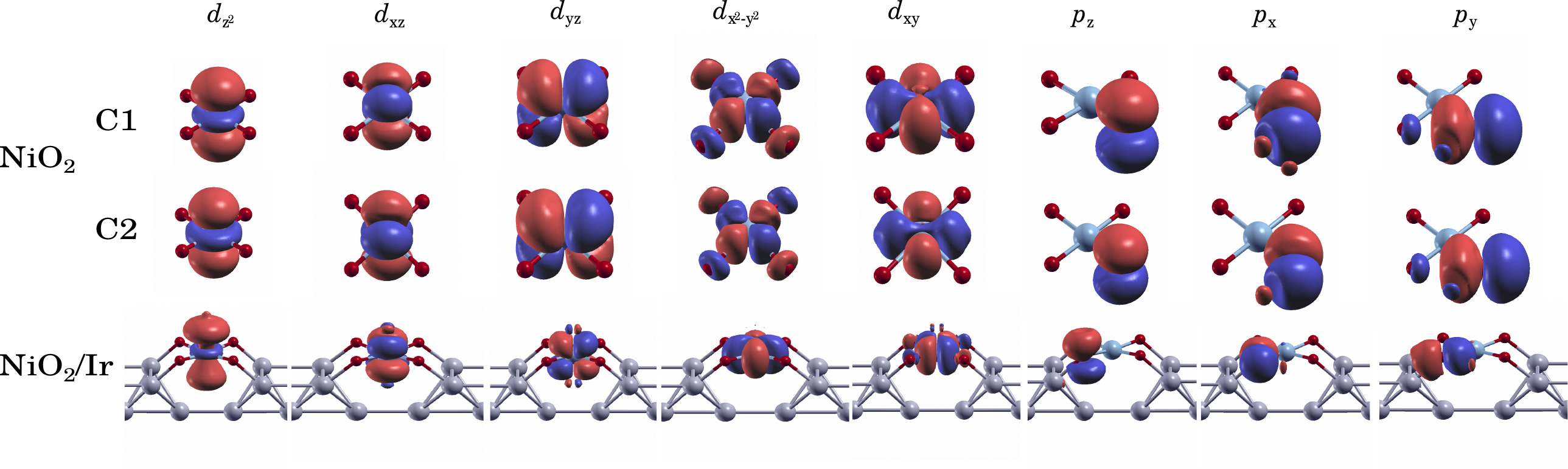}}
\caption{\textbf{Maximally localized Wannier Functions (MLWF) of NiO$_2$ chains.}
Real-space representation of the MLWFs with Ni($d,\downarrow$) and O($p,\downarrow$)
character obtained from the spin-minority electronic wavefunctions of
free-standing (top row) and Ir-supported (bottom row) NiO$_2$. 
Both electronic configurations of the unsupported chain are included.}
\label{fig:mlwf_nio2}
\end{figure}

\begin{figure}[tb!]
\centerline{\includegraphics[width=1\columnwidth]{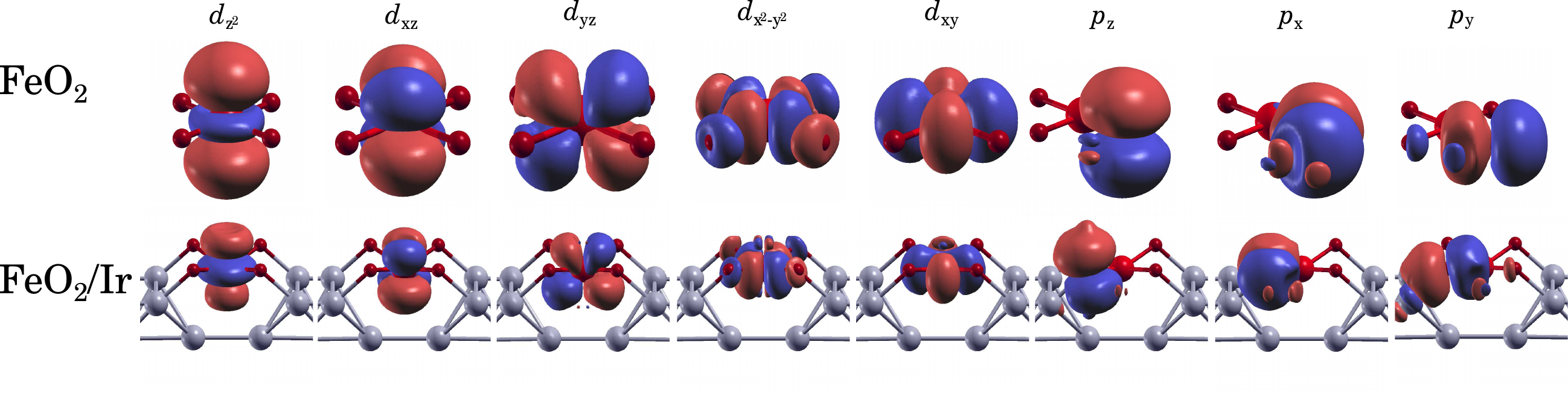}}
\caption{\textbf{MLWFs of FeO$_2$ chains.}
Same information as in Fig.~\ref{fig:mlwf_nio2} for the FeO$_2$ case.}
\label{fig:mlwf_feo2}
\end{figure}

\clearpage

\bibliography{uchains}

\makeatletter\@input{aux4_uchains_v7.tex}\makeatother